\documentclass[aps,twocolumn,showpacs,showkeys,amsmath,amssymb,superscriptaddress,floatfix,nofootinbib]{revtex4}

\usepackage{graphicx}
\usepackage{bm} 
\usepackage{subfigure}

\makeatletter
\newcommand\erfc{\mathop{\operator@font erfc}\nolimits}
\def\slashchar#1{\setbox0=\hbox{$#1$}
   \dimen0=\wd0 \setbox1=\hbox{/} \dimen1=\wd1
   \ifdim\dimen0>\dimen1 \rlap{\hbox to \dimen0{\hfil/\hfil}} #1
   \else  \rlap{\hbox to \dimen1{\hfil$#1$\hfil}} / \fi}

\begin{document}
 
\title{Highly-anisotropic hydrodynamics in 3+1 space-time dimensions \footnote{This work was supported in part by the Polish Ministry of Science and Higher Education under Grants No. N N202 263438 and No. N N202 288638.}}

\author{Radoslaw Ryblewski} 
\email{Radoslaw.Ryblewski@ifj.edu.pl}
\affiliation{The H. Niewodnicza\'nski Institute of Nuclear Physics, Polish Academy of Sciences, PL-31342 Krak\'ow, Poland}

\author{Wojciech Florkowski} 
\email{Wojciech.Florkowski@ifj.edu.pl}
\affiliation{Institute of Physics, Jan Kochanowski University, PL-25406~Kielce, Poland} 
\affiliation{The H. Niewodnicza\'nski Institute of Nuclear Physics, Polish Academy of Sciences, PL-31342 Krak\'ow, Poland}

\date{April 10, 2012}

\begin{abstract}
Recently formulated model of highly-anisotropic and strongly dissipative hydrodynamics is used in 3+1 dimensions to study behavior of matter produced in ultra-relativistic heavy-ion collisions. We search for possible effects of the initial high anisotropy of pressure on the final soft-hadronic observables. We find that by appropriate adjustment of the initial energy density and/or the initial pseudorapidity distributions, the effects of the initial anisotropy of pressure may be easily compensated and the final hadronic observables become insensitive to early dynamics. Our results indicate that the early thermalization assumption is not necessary to describe hadronic data, in particular, to reproduce the measured elliptic flow $v_2$. The complete thermalization of matter (local equilibration) may take place only at the times of about 1--2 fm/c, in agreement with the results of microscopic models.
\end{abstract}



\maketitle 

\section{Introduction}
\label{sect:intro}

The soft-hadronic data collected in ultra-relativistic heavy-ion collisions at RHIC are described very well by the perfect-fluid hydrodynamics or by viscous hydrodynamics with a small viscosity to entropy ratio (for recent reviews see, for example, \cite{Florkowski:2010zz} and \cite{Heinz:2009xj}, respectively).  In particular, a very good description of the soft region of the transverse-momentum spectra of hadrons has been achieved in those frameworks. The shape of the spectra as well as their azimuthal dependence are interpreted as the evidence for the {\it radial} and {\it elliptic} flow of matter. One of the common ingredients of the successful hydrodynamic calculations is {\it an early starting time assumed for the hydrodynamic evolution}, quite often being a fraction of a fermi. Since the initial starting time is identified with the thermalization time, the question arises whether such incredibly fast thermalization of matter can be explained by microscopic models of the very early stages of heavy-ion collisions \cite{Florkowski:2010zz}.

Fast equilibration and perfect-fluidity are naturally explained by the concept that the produced matter is a {\it strongly coupled quark-gluon plasma} (sQGP) \cite{Shuryak:2004cy,Gyulassy:2004zy}. On the other hand, the asymptotic freedom property of QCD suggests that the quark-gluon plasma should behave as a {\it weakly interacting gas of quasiparticles} (wQGP). It is difficult to find convincing explanations for sufficiently fast equilibration of such a weakly interacting system. In particular, at the very early stages of heavy-ion collisions, the momentum distribution of produced partons is expected to be highly anisotropic, which leads to the initial high anisotropy of pressure. There are arguments that the momentum-space anisotropies may persist to several fermis \cite{Martinez:2009mf}. 

The presence of a highly-anisotropic phase suggests that the standard approach to heavy-ion collisions, based on the perfect-fluid or viscous hydrodynamics, may be inappropriate as it assumes that the system is always close to local equilibrium, i.e., its momentum distribution is very close to being isotropic. These observations initiated development of effective, hydrodynamics-like models which relax the assumption of the early thermalization. Typically, this is achieved by imposing existence of a short initial pre-equilibrium phase followed by the standard hydrodynamic evolution.  An example of such a model is presented in Ref. \cite{Broniowski:2008qk}, where the authors assume that the perfect-fluid stage is preceded by free streaming of partons, see also \cite{Sinyukov:2006dw,Gyulassy:2007zz,Qin:2010pf}. Another example is a model discussed in Ref. \cite{Ryblewski:2010tn}, where the initial stage consists of partons with thermalized transverse degrees of freedom only \cite{Bialas:2007gn}.

Recently, we have formulated a framework of highly-anisotropic and strongly-dissipative hydrodynamics (\texttt{ADHYDRO}) \cite{Florkowski:2010cf,Ryblewski:2010ch,Ryblewski:2010bs,Ryblewski:2011aq,Florkowski:2011jg}. This model interpolates between a  highly-anisotropic  initial state  (where the longitudinal and transverse pressures may be substantially different from each other) and the regime described by the perfect-fluid hydrodynamics. It may be used to analyze the effects of early anisotropic pressure on the subsequent evolution of matter. A similar framework has been developed independently by Martinez and Strickland \cite{Martinez:2010sc,Martinez:2010sd,Martinez:2012tu}, see also \cite{Strickland:2011mw,Strickland:2011aa}. The main difference is that our approach is based on the entropy source that has been postulated to agree with the basic physical constraints, while in the Martinez-Strickland approach the equations of motion have been derived by taking the moments of the Boltzmann equations treated in the relaxation-time approximation. Another difference is that we use the constant relaxation time defining the approach to local equilibrium, while Martinez and Strickland use the relaxation time that is proportional to the typical momentum scale of the particles of the system. The similarities and differences between the two frameworks have been analyzed in Ref.~\cite{Ryblewski:2010ch}. The connection to the Israel-Stewart theory has been discussed in Refs.~\cite{Martinez:2010sc,Florkowski:2010cf,Florkowski:2011jg}.

The concept of anisotropic hydrodynamics has been recently approached from the point of view of strongly coupled gauge theories \cite{Witaszczyk:2008zz,Janik:2008tc,Heller:2011ju,Heller:2012je,Mateos:2011ix,Mateos:2011tv,Heller:2012km,Chernicoff:2012iq,Chernicoff:2012gu,Giataganas:2012zy,Gahramanov:2012wz}. It would be very much interesting to combine the results of such studies with our phenomenological approach. In particular, it would be of great interest to gain some hints about the entropy source term appearing in our framework from the strong-coupling approach. 

In this paper we use our model for the first time in 3+1 space-time dimensions (\texttt{(3+1)D ADHYDRO}). We analyze possible effects of the initial high anisotropy of pressure on the final soft-hadronic observables. We find that by appropriate modification of the initial energy density and/or the initial pseudorapidity distribution of matter, the effects of the initial anisotropy of pressure may be easily compensated: the same final hadronic observables are reproduced for different initial conditions, including those with high pressure anisotropy. Consequently, our results indicate that the early thermalization assumption is not necessary to describe the hadronic data. The complete thermalization of matter (local equilibration) may take place only at the times of about 1--2 fm/c, in agreement with the results of microscopic models. The preliminary results discussed here have been presented at the SQM2011 conference \cite{Ryblewski:2011ye}.

\section{Space-time dynamics of highly-anisotropic systems}
\label{sect:hydro-general}
%
In our framework, the space-time dynamics of a highly-anisotropic system is governed by the two equations,
\begin{eqnarray}
\partial_\mu T^{\mu \nu} &=& 0, \label{enmomcon} \\
\partial_\mu S^{\mu} &=& \Sigma. \label{engrow}
\end{eqnarray}
The first equation has the standard form expressing the energy and momentum conservation law.  The second equation describes the entropy production determined by the entropy source $\Sigma$. In the standard perfect-fluid approach, the right-hand-side of Eq.~(\ref{engrow}) vanishes and Eq.~(\ref{engrow}) becomes the condition of adiabaticity of the flow. 

The energy momentum tensor and the entropy flow used by us in (\ref{enmomcon}) and (\ref{engrow}) have the forms
\begin{eqnarray}
T^{\mu \nu} = \left( \varepsilon  + P_{\perp}\right) U^{\mu}U^{\nu} - P_{\perp} \, g^{\mu\nu} - (P_{\perp} - P_{\parallel}) V^{\mu}V^{\nu}
\label{Taniso}
\end{eqnarray}
and
\begin{equation}
S^{\mu} = \sigma \, U^\mu .
\label{Saniso}
\end{equation}
Equations (\ref{Taniso}) and (\ref{Saniso}) contain energy density $\varepsilon$, transverse pressure $P_\perp$, longitudinal pressure $P_\parallel$, and entropy density $\sigma$~\footnote{The transverse and longitudinal directions are always defined with respect to the beam axis.}. Below, these parameters will be named and treated as thermodynamic parameters, although, strictly speaking, they do not describe the equilibrium state. Only if the local equilibrium is reached by the system, they gain standard thermodynamic interpretation and become interconnected via the equation of state (EOS).

We note that in the perfect-fluid hydrodynamics Eq.~(\ref{engrow}), with $\Sigma=0$, follows directly from Eq.~(\ref{enmomcon}) and from the appropriate form of the energy-momentum tensor, hence, it cannot be used as an independent condition. In our approach, the transverse and longitudinal pressures are different, and Eq.~(\ref{engrow}) should be treated as an ansatz. The specific form of the entropy source, for example, in the form $\Sigma=\Sigma(P_\perp,P_\parallel)$, allows us to close the system of equations and determines the fluid dynamics.

The four-vector $U^\mu$ describes the fluid four-velocity
\begin{equation}
U^\mu = \gamma (1, v_x, v_y, v_z), \quad \gamma = (1-v^2)^{-1/2}.
\label{U}
\end{equation}
In contrast to standard perfect-fluid hydrodynamics, we introduce a new four-vector $V^\mu$. Its appearance is connected with the special role played by the beam axis ($z$-axis) and it is defined as
\begin{equation}
V^\mu = \gamma_z (v_z, 0, 0, 1), \quad \gamma_z = (1-v_z^2)^{-1/2}.
\label{V}
\end{equation}
The four-vectors $U^\mu$ and $V^\mu$ satisfy simple normalization conditions
\begin{eqnarray}
U^2 = 1, \quad V^2 = -1, \quad U \cdot V = 0.
\label{UVnorm}
\end{eqnarray}
In the local rest frame (LRF) of the fluid element the four-vectors $U^\mu$ and $V^\mu$ have simple forms
\begin{eqnarray}
 U^\mu = (1,0,0,0), \quad V^\mu = (0,0,0,1).
 \label{UVLRF}
\end{eqnarray}
Since we have $T^{\mu \nu} U_\nu = \varepsilon U^\mu$, the four-vector $U^\mu$ corresponds to Landau's definition of the hydrodynamic flow.

The projections of Eq.~(\ref{enmomcon}) on the four-vectors $U_\nu$ and $V_\nu$ yield 
\begin{eqnarray}
U^\mu \partial_\mu \varepsilon &=& - \left( \varepsilon+P_\perp \right) \partial_\mu U^\mu 
+ \left( P_\perp-P_\parallel \right) U_\nu  V^\mu \partial_\mu V^\nu, \nonumber \\ \label{enmomconU} \\
V^\mu \partial_\mu P_\parallel &=& - \left( P_\parallel-P_\perp \right) \partial_\mu V^\mu 
+ \left( \varepsilon+ P_\perp \right) V_\nu  U^\mu \partial_\mu U^\nu. \nonumber \\ \label{enmomconV}
\end{eqnarray}
Equation (\ref{enmomconU}) expresses the energy conservation, whereas Eq.~(\ref{enmomconV}) describes the conservation of the longitudinal momentum.

\section{Generalized equation of state}
\label{sect:aniso-eos}

The equations of perfect-fluid hydrodynamics form a closed system of equations only if they are supplemented with EOS that specifies thermodynamic quantities as functions of {\it one} parameter (if the baryon-free matter is considered). Similarly, our framework requires that all thermodynamic quantities should be expressed by {\it two} parameters (again, if the baryon-free matter is considered). Such relations play a role of the {\it generalized equation of state} and allow us to close the system of dynamic equations.
  
If one considers a system of partons described by the anisotropic phase-space distribution function obtained by squeezing (or stretching) the longitudinal momentum in the classical Boltzmann distribution (the Romatschke-Strickland ansatz \cite{Romatschke:2003ms}), the thermodynamic parameters have the form
\begin{eqnarray}
\varepsilon (x,\sigma)&=&  \varepsilon_{\rm id}(\sigma) r(x), \label{epsilon2a}   \nonumber \\
P_\perp (x,\sigma)&=&  P_{\rm id}(\sigma) \left[r(x) + 3 x r^\prime(x) \right],  \label{GEOS}  \\ 
P_\parallel (x,\sigma)&=&  P_{\rm id}(\sigma) \left[r(x) - 6 x r^\prime(x) \right], \nonumber 
\end{eqnarray}
where
\begin{equation}
r(x) = \frac{x^{-\frac{1}{3}}}{2} \left[ 1 + \frac{x \arctan\sqrt{x-1}}{\sqrt{x-1}}\right]
\label{RB}
\end{equation}
and $r^\prime(x)$ denotes the derivative $dr/dx$. In Eqs. (\ref{GEOS}) $\sigma$ is the entropy density and $x$ defines the anisotropy of the underlying momentum distribution. To a good approximation one has $P_\parallel/P_\perp = x^{-3/4}$. The functions $\varepsilon_{\rm id}(\sigma)$ and $P_{\rm id}(\sigma)$ are {\it equilibrium expressions} for the energy density and pressure, however, calculated typically for the {\it non-equilibrium} entropy density $\sigma$~\footnote{We note that for massless classical particles in equilibrium the following relations are valid: $\varepsilon_{\rm id} = 3 g_0 T^4/\pi^2$, $P_{\rm id} = g_0 T^4/\pi^2$, and $\sigma_{\rm id} = 4 g_0 T^3/\pi^2$, where $g_0$ is the number of internal degrees of freedom \cite{Florkowski:2010zz}. This implies the functional relation $\varepsilon_{\rm id}(\sigma) = 3 g_0/\pi^2 \,(\pi^2 \sigma/(4 g_0))^{4/3}$ which is used in (\ref{GEOS}).}. 

The generalized equation of state (\ref{epsilon2a}) is very likely to be realized at the very early stages of the collisions, when the produced system is highly anisotropic and described by the distribution functions discussed in Ref.~\cite{Romatschke:2003ms}. On the other hand, as the system expands, the interactions between its constituents thermalize it, so the system's thermodynamic variables approach the realistic QCD EOS and $x$ tends to unity (in this case we have $r(1)=1$ and $r'(1)=0$). 

The realistic EOS for strongly interacting matter has been constructed in Ref. \cite{Chojnacki:2007jc}. At low temperatures it is consistent with the hadron-gas model with all experimentally identified resonances. On the other hand, at high temperatures it coincides with the lattice simulations of QCD (LQCD). The newest development of the LQCD EOS presented in Ref.~\cite{Borsanyi:2010cj} gives EOS which leads to practically the same results as \cite{Chojnacki:2007jc} when used in hydrodynamic calculations. 

The two limiting cases discussed above suggest that one may use the generalized equation of state of the form
\begin{eqnarray}
\varepsilon (x,\sigma)&=&  \varepsilon_{\rm qgp}(\sigma) r(x),  \nonumber  \\ 
P_\perp (x,\sigma)&=&  P_{\rm qgp}(\sigma) \left[r(x) + 3 x r^\prime(x) \right], \label{GEOS2}   \\ 
P_\parallel (x,\sigma)&=&  P_{\rm qgp}(\sigma) \left[r(x) - 6 x r^\prime(x) \right]. \nonumber 
\end{eqnarray}
Here, the equation of state of the ideal relativistic gas, defined by the functions $\varepsilon_{\rm id}(\sigma)$ and $P_{\rm id}(\sigma)$ has been replaced by the formulas characterizing the realistic equation of state for vanishing baryon chemical potential.

We have to stress that there is no microscopic explanation for the form of Eqs. (\ref{GEOS2}). Despite of this fact, Eqs.~(\ref{GEOS2}) have many attractive features. Firstly, as discussed above, they reproduce properly the two important limits and they interpolate between them in a continuous way. Moreover there are no additional assumptions needed (such as, e.g., the Landau matching conditions) in order to describe the change from one thermodynamic regime to another one. Additionally, Eqs. (\ref{GEOS2}) naturally include the phase transition from the quark-gluon plasma to the hadron gas. In this way Eqs. (\ref{GEOS2}) are able to describe different stages of evolution of matter in heavy-ion collisions in a uniform way.

\section{Entropy source}
\label{sect:hydro-entropy}

The entropy source $\Sigma$ describes the entropy growth due to equilibration of pressures in the system. In the following we treat it as a function of $\sigma$ and $x$ rather than as a function of the two pressures. As we have already mentioned above, Eqs.~(\ref{enmomcon}) and  (\ref{engrow}) may be solved only if the function $\Sigma(\sigma,x)$ is specified. The functional form $\Sigma(\sigma, x)$ must be delivered as the external input for the anisotropic hydrodynamics. 

The simplest ansatz for $\Sigma(\sigma,x)$ satisfying general physical assumptions may be proposed in the following form
\begin{equation}
\Sigma = \frac{(1-\sqrt{x})^{2}}{\sqrt{x}}\frac{\sigma}{\tau_{\rm eq}},
\label{en1}
\end{equation}
where $\tau_{\rm eq}$ is a time-scale parameter introduced to control the rate of the processes leading to equilibration of the system. The expression on the right-hand-side of Eq.~(\ref{en1}) has several appealing features. First of all it is positive, as expected on the grounds of the second law of thermodynamics. Moreover, it has a correct dimension, since $\Sigma$ is proportional to the entropy density $\sigma$. In the natural way, the entropy source $\Sigma$ defined by (\ref{en1}) vanishes in equilibrium, where $x=1.0$, thus Eqs.~(\ref{enmomcon}) and (\ref{engrow}) can be reduced to the equations of perfect fluid hydrodynamics in the limit $x \to 1$. Finally, for small deviations from equilibrium, where $|x-1| \ll 1$, we find
\begin{equation}
\Sigma (x) \approx \frac{(x-1)^{2}}{4 \tau_{\rm eq}} \sigma.
\label{en1exp}
\end{equation}
The quadratic dependence of the entropy source displayed in (\ref{en1exp}) is characteristic for the 2nd order viscous hydrodynamics. For more details about connections of our approach with the Israel-Stewart theory see Refs.~\cite{Martinez:2010sc,Florkowski:2010cf,Florkowski:2011jg}.

\section{3+1 space-time expansion}
\label{sect:hydro-full}

In the general case, where matter expands in the longitudinal and transverse directions without any symmetry constraints, we may use the following parametrization of the four-velocity of the fluid $U^\mu$ and the four-vector $V^{\mu}$ 
\begin{eqnarray}
U^\mu &=& (u_0 \cosh \vartheta, u_x, u_y, u_0 \sinh \vartheta), \label{U3+1} \\
V^\mu &=& (	 \sinh \vartheta, 0, 0,  \cosh \vartheta). \label{V3+1}
\end{eqnarray}
In Eqs.~(\ref{U3+1}) and (\ref{V3+1}) we have introduced the two components of the transverse four-velocity of the fluid, $u_x$ and $u_y$, and the longitudinal fluid rapidity $\vartheta$. Using the normalization condition (\ref{UVnorm}) we also find  
\begin{eqnarray}
u_0 = \sqrt{1+u_\perp^2}, \quad \quad u_\perp = \sqrt{u_x^2 + u_y^2}. \label{u0}
\end{eqnarray}
If the parametrizations (\ref{U3+1}) and (\ref{V3+1}) are used, the differential operators appearing in (\ref{enmomconU}) and (\ref{enmomconV}) take the following form
\begin{eqnarray}
U^\mu \partial_\mu &=& {\bf u}_\perp \cdot {\bf \nabla}_\perp + u_0 L_1, \nonumber \\
V^\mu \partial_\mu &=&  L_2, \nonumber \\
\partial_\mu U^\mu &=& {\bf \nabla}_\perp \cdot {\bf u}_\perp + L_1 u_0 + u_0 L_2 \vartheta, \nonumber \\
\partial_\mu V^\mu &=&  L_1 \vartheta, \label{op1} \\
U_\nu V^\mu \partial_\mu V^\nu &=& u_0 L_2 \vartheta, \nonumber \\
V_\nu U^\mu \partial_\mu U^\nu &=& - u_0 \left( {\bf u}_\perp \cdot 
{\bf \nabla}_\perp + u_0 L_1  \right) \vartheta \nonumber, 
\end{eqnarray}
where the two linear differential operators $L_1$ and $L_2$ are defined by the expressions
\begin{eqnarray}
L_1 &=& \cosh Y \partial_\tau - \sinh Y \frac{\partial_\eta}{\tau}, \label{op21}\\
L_2 &=& - \sinh Y \partial_\tau + \cosh Y \frac{\partial_\eta}{\tau}.\label{op22}
\end{eqnarray}
In Eqs.~(\ref{op21}) and (\ref{op22}) we have introduced the space-time rapidity $\eta$ and the proper time $\tau$,
\begin{eqnarray}
\eta &=& \frac{1}{2} \ln \frac{t+z}{t-z}, \label{eta}\\
\tau &=& \sqrt{t^2 - z^2}. \label{tau}
\end{eqnarray}
The difference between the longitudinal fluid rapidity $\vartheta$ and the space-time rapidity $\eta$ is defined as 
\begin{equation}
Y = \eta - \vartheta.
\label{Yetatheta}
\end{equation}
We use the boldface notation for two-dimensional vectors in the transverse plane and write: ${\bf u}_\perp = (u_x,u_y)$ and 
\mbox{${\bf \nabla}_{\perp}=(\partial_x,\partial_y )$}. 

Besides Eqs.~(\ref{enmomconU}) and (\ref{enmomconV}), the two additional equations describing transverse dynamics of the system are needed. They can be chosen, for example, as the two linear combinations: $U_{1} \partial_{\mu} T^{\mu 1} +   U_{2} \partial_{\mu} T^{\mu 2} = 0$ and $U_{2} \partial_{\mu} T^{\mu 1} -  U_{1} \partial_{\mu} T^{\mu 2} = 0$. In this case we find
\begin{eqnarray}
& & \mathcal{D} u_{\perp} = - \frac{u_{\perp}}{\varepsilon + P_{\perp}} 
\left[ \frac{ {\bf u}_\perp \cdot {\bf \nabla}_\perp P_{\perp}}{u_\perp^2}  
+ \mathcal{D} P_{\perp} \right. \label{HydEqEuler1} \\
& & \left. \hspace{2.75cm} + (P_{\perp} - P_{\parallel}) U_\nu V^\mu \partial_\mu V^\nu \right],  
\nonumber \\
& & \mathcal{D} \left( \frac{u_x}{u_y} \right) = \frac{1}{u_y^2 (\varepsilon + P_{\perp})} 
\left( u_x \partial_y - u_y \partial_x \right)P_{\perp}.
\label{HydEqEuler2}
\end{eqnarray}
Here $\mathcal{D} = U^\mu \partial_\mu$ is the total time derivative  and $\theta = \partial_\mu U^\mu$ is the volume expansion rate. Using this notation, the entropy production equation, see Eq.~(\ref{engrow}), can be put in the compact form
\begin{equation}
\mathcal{D} \sigma + \sigma \theta = \Sigma.
\label{entrprod}
\end{equation}

Equation (\ref{entrprod}) together with Eqs.~(\ref{enmomconU}), (\ref{enmomconV}), (\ref{HydEqEuler1}) and (\ref{HydEqEuler2}) form a set of hydrodynamic equations describing the (3+1)D dynamics of the anisotropic fluid --- the \texttt{(3+1)D ADHYDRO} approach. We note that these equations have the same structure as those used in the (2+1)D version of our model \cite{Ryblewski:2011aq}, however, differential operators are defined now by more complex expressions given by Eqs.~(\ref{op1}).

If the generalized equation of state $\varepsilon(\sigma,x)$ and the entropy source $\Sigma(\sigma,x)$ are specified, the \texttt{(3+1)D ADHYDRO} equations form a closed system of five equations for five unknown functions: two components of the fluid velocity $u_x$ and $u_y$, the longitudinal rapidity of the fluid $\vartheta$, the non-equilibrium entropy density $\sigma$ and the anisotropy parameter $x$. These functions depend on  transverse coordinates $x,y$, the space-time rapidity $\eta$, and the proper time $\tau$. One has to solve the \texttt{(3+1)D ADHYDRO} equations numerically for a given initial condition specified at a certain initial time $\tau_0$.

\section{Initial conditions and freeze-out}
\label{section:tini}

\subsection{Tilted source}

In this paper, we analyze Au+Au collisions at the highest beam energy studied at RHIC, i.e., at \mbox{$\sqrt{s_{\rm NN}} = 200$~GeV}. For general non-boost-invariant cases, the initial conditions for \texttt{(3+1)D ADHYDRO} are defined by five functions: $\sigma(\tau_0,\eta,{\bf x}_\perp)$, $x(\tau_0,\eta,{\bf x}_\perp)$, $u_x(\tau_0,\eta,{\bf x}_\perp)$, $u_y(\tau_0,\eta,{\bf x}_\perp)$, and $\vartheta(\tau_0,\eta,{\bf x}_\perp)$. Throughout the paper we use $\tau_0$=0.25 fm.

The initial entropy density profile has the form
\begin{equation}
 \sigma_0(\eta,{\bf x}_\perp) = \sigma(\tau_0,\eta,{\bf x}_\perp) = \varepsilon_{\rm gqp}^{-1} 
\left[ \varepsilon_{\rm i} \, \tilde{\rho}(b,\eta,{\bf x}_\perp) \right],
\label{sig2}
\end{equation}
where $\varepsilon_{\rm gqp}^{-1}$ is the inverse function to that used in Eq.~(\ref{GEOS2}), $b$ is the impact parameter, and $\tilde{\rho}(b,\eta,{\bf x}_\perp)$ is the {\it normalized density of sources}, 
\begin{equation}
 \tilde{\rho}(b,\eta,{\bf x}_\perp) = \frac{\rho(b,\eta,{\bf x}_\perp)}{\rho(0,0,0)}.
 \label{dsourcest}
\end{equation}

The quantity $\varepsilon_{\rm i}$ is the initial energy density at the center of the system created in the most central collisions. Its value is fixed by the measured multiplicity, separately for different physical scenarios considered in this paper. These scenarios correspond to different initial anisotropy profiles and/or different values of the time-scale parameter $\tau_{\rm eq}$. For the cases where the initial anisotropy is constant in space, the appropriate values of $\varepsilon_{\rm i}$ are given in Table~\ref{table:multiplicity}. 

The density of sources describing the tilted source, introduced for the first time by Bozek and Wyskiel in Ref.~\cite{Bozek:2010bi}, is given by the expression
\begin{eqnarray}
\rho(b,\eta,{\bf x}_\perp) &=&  (1-\kappa)\left[
\rho_{\rm W}^{+} \left(b,{\bf x}_\perp \right) f^{+} \left(\eta\right) \right. \label{dsources2} \\
& & \hspace{0.15cm} + \left.
\rho_{\rm W}^{-} \left(b,{\bf x}_\perp \right) f^{-} \left(\eta\right) 
\right] + \kappa \rho_{\rm B} \left(b,{\bf x}_\perp \right) f \left(\eta\right), \nonumber 
\end{eqnarray}
where $f \left(\eta\right)$ is the initial longitudinal profile
\begin{equation}
f \left(\eta\right)= \exp 
\left[   
- \frac{(\eta - \Delta\eta)^2}{2 \sigma_\eta^2} \theta(|\eta|-\Delta\eta)
\right].
\label{longitprof}
\end{equation}
The half-width of the central plateau, $\Delta\eta$, and the  half-width of Gaussian tails,  $\sigma_\eta$, are fitted to reproduce the RHIC pseudorapidity distributions. 

The terms in the square brackets on the right-hand-side of Eq.~(\ref{dsources2}) introduce contributions to the density of sources from the forward ($+$) or backward-moving ($-$) wounded nucleons \cite{Bialas:1976ed}. This formula assumes a preferred emission from the wounded nucleons along the direction of their motion \cite{Bialas:2004su}. This contribution is taken with the weight $(1-\kappa)$, where we set $\kappa=0.14$ \cite{Chojnacki:2007rq}. The last term on the right-hand-side of Eq.~(\ref{dsources2}) gives a symmetric contribution from the binary collisions (taken with the weight $\kappa$). 

The initial density of sources produced by a single forward- or backward-moving wounded nucleon with rapidity $\mathrm{y}_b = \ln (\sqrt{s_{\rm NN}}/m_{\rm N})$, where $m_{\rm N}$ is the nucleon mass, is proportional to
\begin{equation}
f^{+(-)} \left(\eta\right)= f \left(\eta\right) f_{F(B)} \left(\eta\right),
\label{fplusminus}
\end{equation}
where
\begin{equation}
f_F(\eta)=
\begin{cases} 0 & \eta< -\eta_m \\
\frac{\eta+\eta_m}{2\eta_m}   & -\eta_m \le \eta \le \eta_m \\
1 & \eta_m<\eta
\end{cases}, \quad \quad
f_B(\eta)=f_F(-\eta).
\label{fForBack}
\end{equation}
The range of rapidity correlations, $\eta_m =\mathrm{y}_b-\eta_s \simeq 3.36$, and the shift in the rapidity, $\eta_s = 2$, are the same as those used in the (3+1)D hydrodynamic calculation described in Ref.~\cite{Bozek:2010bi}. 

Similarly to our earlier (2+1)D calculations~\cite{Ryblewski:2011aq}, for the initial anisotropy profile, $x(\tau_0,\eta,{\bf x}_\perp)$, we choose three different values: $x_0=100$, $x_0=1.0$, and $x_0=0.032$~\footnote{We choose $x_0=0.032$ because $r(100) \approx r(0.032)$, hence, for the fixed initial entropy density the cases $x_0=100$ and $x_0=0.032$ have the same energy density but the role of the transverse and longitudinal pressure is exchanged.}. Moreover, we check the possibility of the spatial dependence of the initial anisotropy parameter, where $x_0=x_0( \tilde{\rho}(\eta,{\bf x}_\perp))$. We assume also that there is no transverse flow present initially, therefore, we set $u_x(\tau_0,\eta,{\bf x}_\perp)=0$ and $u_y(\tau_0,\eta,{\bf x}_\perp)=0$, and the initial longitudinal rapidity of the fluid follows the simple Bjorken scaling $\vartheta(\tau_0,\eta,{\bf x}_\perp)=\eta$.

\subsection{Freeze-out}

The physical observables are obtained from the Cooper-Frye formalism used in the Monte-Carlo version implemented in \texttt{ THERMINATOR}~\cite{Kisiel:2005hn,Chojnacki:2011hb}. The details of this procedure have been described in our earlier work \cite{Ryblewski:2011aq}. The freeze-out hypersurface is defined by the freeze-out entropy density \mbox{$\sigma_f = 1.79/\hbox{fm}^3$}, corresponding to the freeze-out temperature \mbox{$T_{\rm f} = 150$ $\mathrm{MeV}$}. In the calculation of the freeze-out hypersurface we neglect the first fermi of the time evolution, i.e., we neglect early emission from the edges of the system. In such early times, the system is highly anisotropic and its hadronization cannot be described by the standard Cooper-Frye method. Moreover, at the early times the system consists of gluons which, we assume, do not contribute to the soft hadronic observables. 

\begin{table}[t]
  \begin{center}
    \begin{small}
      \begin{tabular}{lc@{\hskip 1cm}ccc@{\hskip 1cm}ccc}
      \hline \\ [-1ex]
$x_{\rm 0}$ & 1 & \multicolumn{3}{c@{\hskip 1cm}}{100} & \multicolumn{3}{c}{0.032} \\  [1ex] \hline \\ 
$\tau_{\rm eq}  \mathrm{[fm]}$ & 0.25 & 0.25 & 0.5 & 1.0 & 0.25 & 0.5 & 1.0 \\  [2ex]
$\varepsilon_{\rm i} \left[\frac{\hbox{GeV}}{\hbox{fm}^3}\right]$ & 107.5 & 58.3 & 53.4 & 48.8 & 72.5 & 75.9 & 80.1 \\ \\ \hline
      \end{tabular}
    \end{small}
  \end{center}
  \caption{\small Values of the initial central energy density, $\varepsilon_{\rm i}$, used in all the figures of Section VIII, except for Fig.~\ref{fig:etadistr_RHIC_m}.
}  \label{table:multiplicity}
\end{table}

\begin{figure}[t]
\begin{center}
\subfigure{\includegraphics[angle=0,width=0.4\textwidth]{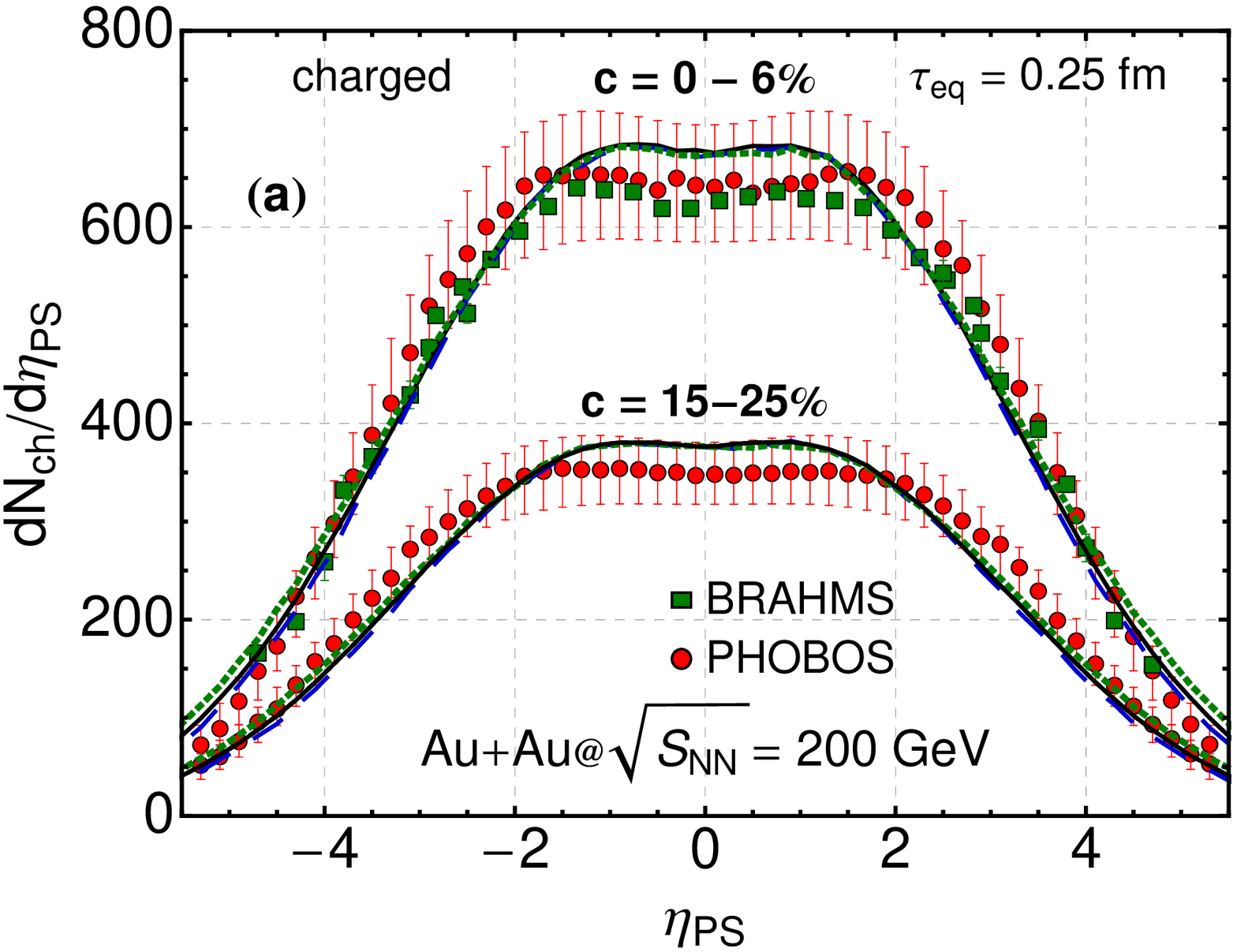}} 
\subfigure{\includegraphics[angle=0,width=0.4\textwidth]{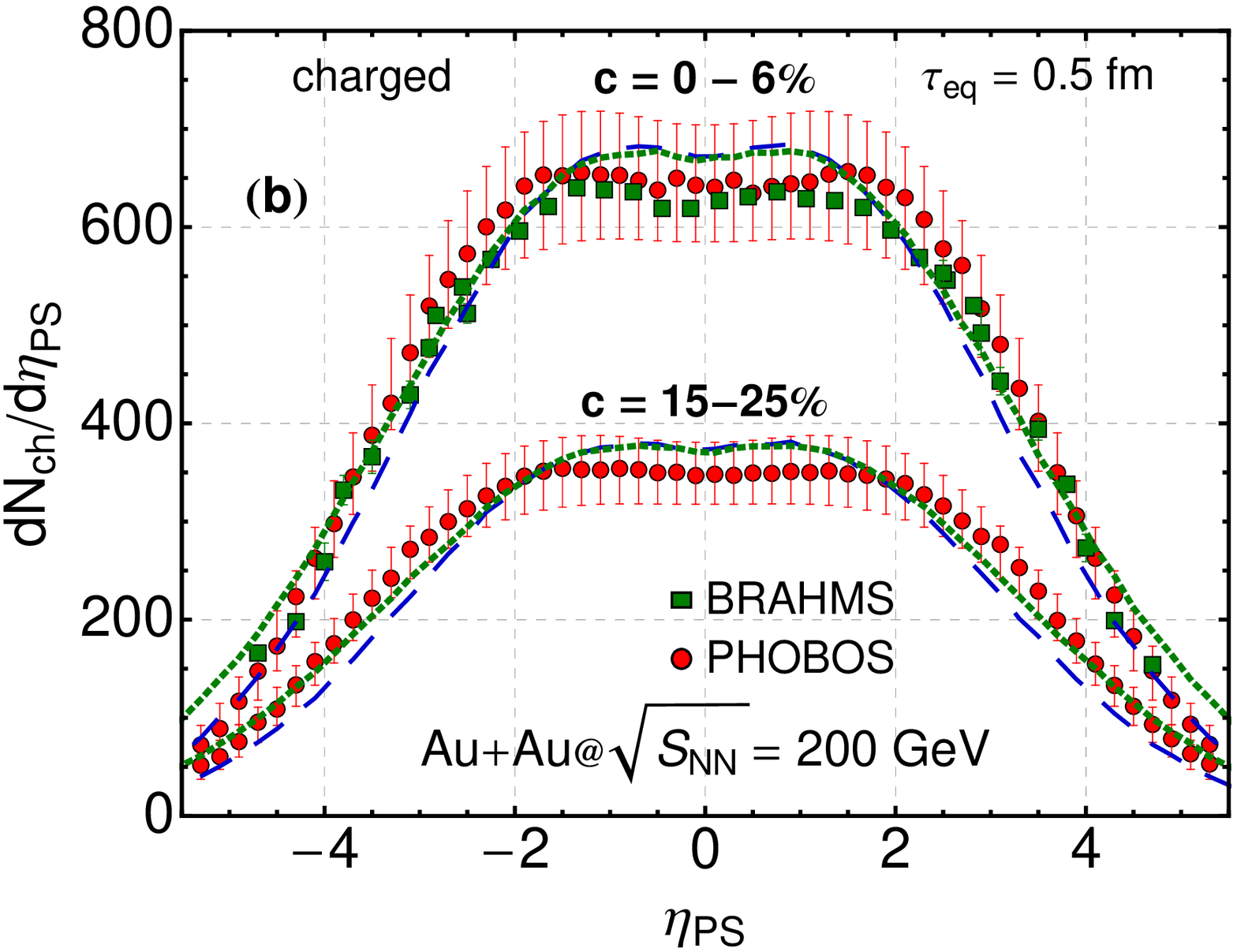}}  
\subfigure{\includegraphics[angle=0,width=0.4\textwidth]{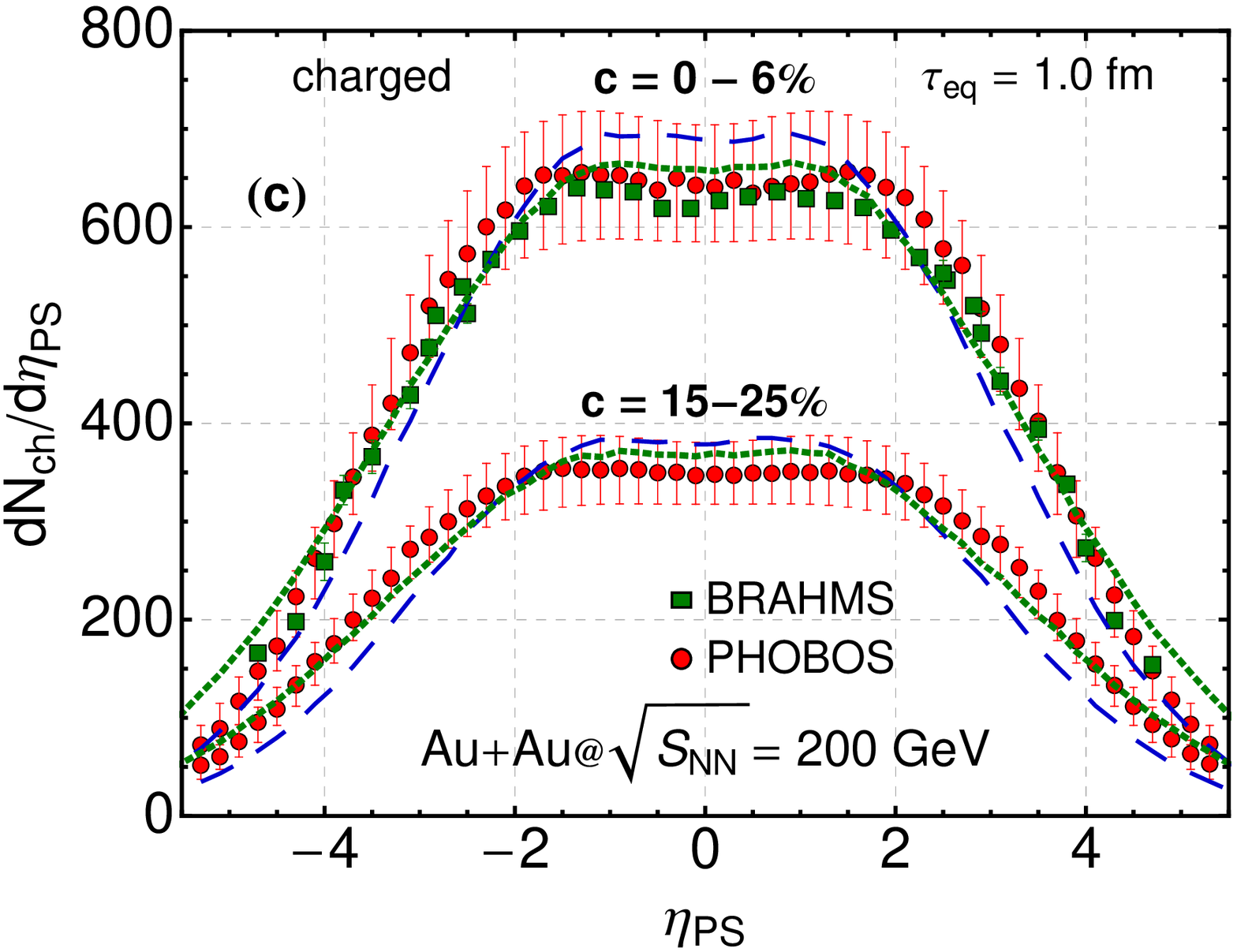}} 
\end{center}
\caption{\small (Color online) The pseudorapidity distribution of charged particles for three values of the time-scale parameter: $\tau_{\rm eq}=0.25$ fm \textbf{(a)}, $\tau_{\rm eq}=0.5$ fm \textbf{(b)}, $\tau_{\rm eq}=1.0$ fm \textbf{(c)}. The results are obtained for three values of the initial momentum anisotropy: $x_{\rm 0}=100$ (dashed blue lines), $x_{\rm 0}=1.0$ (solid black lines) and $x_{\rm 0}=0.032$ (dotted green lines) and the two centrality classes: $c=0-6$\% ($b=2.48$ fm) and $c=15-25$\% ($b=6.4$ fm). The results are compared to experimental data from PHOBOS \cite{Back:2002wb} (red dots) and BRAHMS \cite{Bearden:2001qq} (green squares). Vertical bars for PHOBOS data denote the systematic errors.}
\label{fig:etadistr_RHIC}
\end{figure}

\begin{figure}[t]
\begin{center}
\subfigure{\includegraphics[angle=0,width=0.4\textwidth]{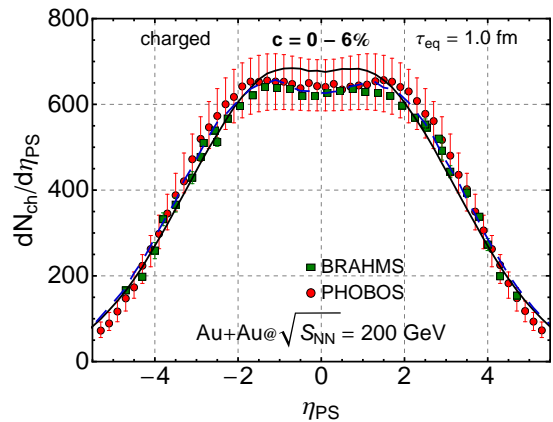}} 
\end{center}
\caption{\small (Color online) The pseudorapidity distribution of charged particles for $x_{\rm 0}=1.0$ (solid black line) and $x_{\rm 0}=100$ (dashed blue line). In the latter case the initial conditions were modified to $\varepsilon_{\rm i} = 41.8$  $\mathrm{GeV/fm^3}$ and $\Delta\eta = 1.5$ to obtain the same final total multiplicity. The results are compared to the experimental PHOBOS  \cite{Back:2002wb} (red dots) and BRAHMS \cite{Bearden:2001qq} (green squares) data.}
\label{fig:etadistr_RHIC_m}
\end{figure}

\section{Constant initial momentum anisotropy}
\label{sect:cima}

In this Section we present our main results. They describe the cases where the initial anisotropy $x_{\rm 0}$ is constant in space and takes the values: \textbf{i)} $x_{\rm 0}=100$, \textbf{ii)} $x_{\rm 0}=1.0$, and \textbf{iii)} $x_{\rm 0}=0.032$. To check how our results depend on the duration of the anisotropic stage, we use three different values of the time-scale parameter: $\tau_{\rm eq}=0.25$ fm, $\tau_{\rm eq}=0.5$ fm, and $\tau_{\rm eq}=1.0$ fm. 

The half-width of the central plateau in the initial profile, $\Delta\eta = 1$, and the half-width of gaussian tails,  $\sigma_\eta = 1.3$, are fixed in this Section (except for the situation described in Fig.~\ref{fig:etadistr_RHIC_m}). These parameters have been fitted to reproduce the shape of the pseudorapidity distribution in the case $x_{\rm 0}=1.0$ and $\tau_{\rm eq}=0.25$ fm. 

The initial energy density in the center of the fireball, $\varepsilon_{\rm i}$, is chosen in such a way as to reproduce the total experimental charged particle multiplicity  $N_{\rm ch}^{\rm exp}= 5060 \pm 250$ \cite{Back:2002wb} for the centrality class $c=0-6$\%. In the case $x_{\rm 0}=1.0$ and $\tau_{\rm eq}=0.25$ fm, we use $\varepsilon_{\rm i}=107.5$ $\mathrm{GeV/fm^3}$ which yields $N_{\rm ch}^{\rm theor}= 5020$, see Table~\ref{table:multiplicity}. 

Since entropy grows during the hydrodynamic evolution due to non-zero entropy production source $\Sigma(\sigma, x)$, see Eq.~(\ref{en1}), the initial central energy density must be appropriately renormalized for each pair of parameters $x_0$ and $\tau_{\rm eq}$ to keep $N_{\rm ch}^{\rm theor}$ unchanged. In the case $x_{\rm 0}=100$ the total entropy produced during the hydrodynamic evolution grows with increasing $\tau_{\rm eq}$, while in the case $x_{\rm 0}=0.032$ the total produced entropy decreases with increasing $\tau_{\rm eq}$ (similarly to the one-dimensional behavior discussed in Ref.~\cite{Florkowski:2010cf}). The values of $\varepsilon_{\rm i}$ are summarized in Table~\ref{table:multiplicity}. The centrality dependence is reproduced by applying the scaling (\ref{dsourcest}). 

\begin{figure}[t]
\begin{center}
\subfigure{\includegraphics[angle=0,width=0.35\textwidth]{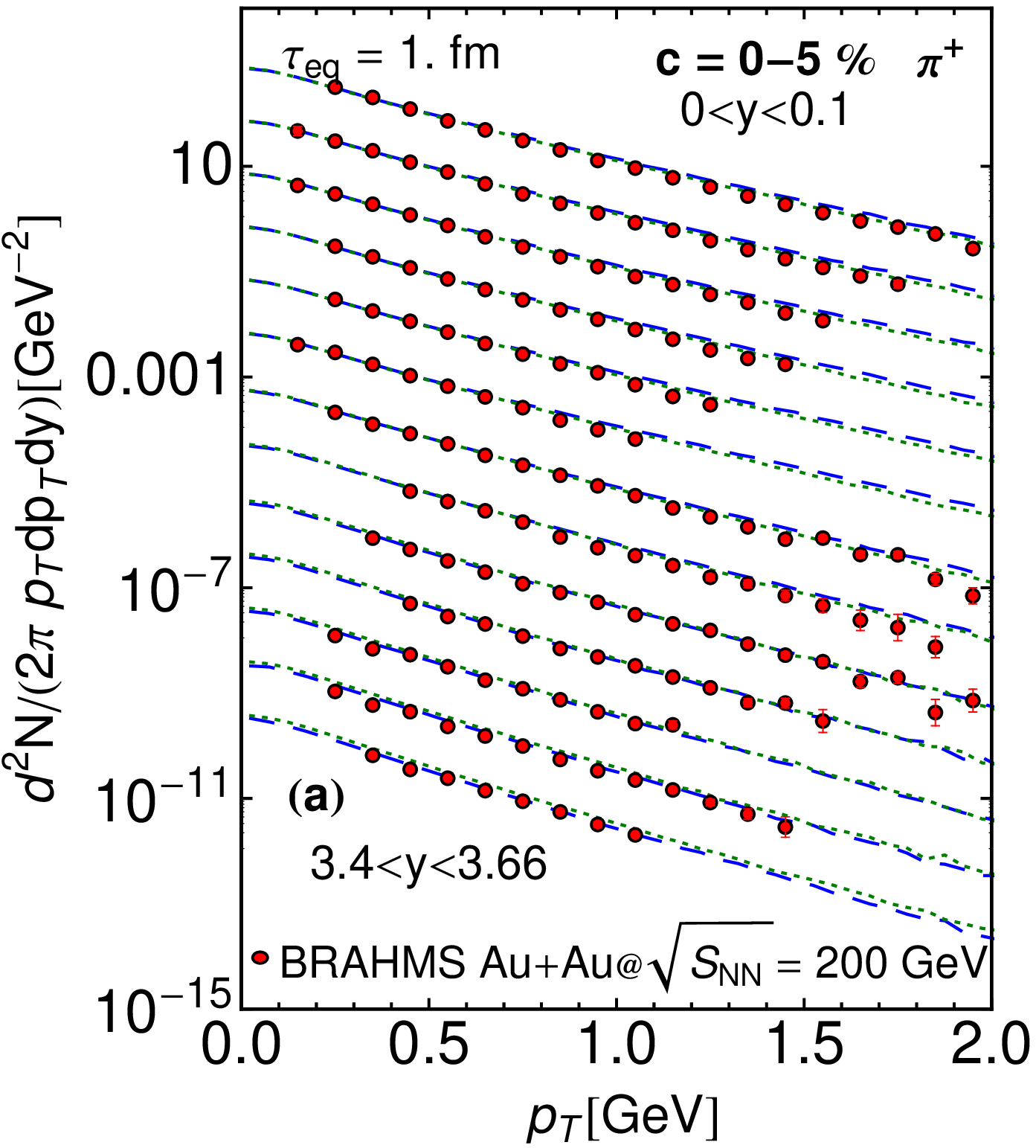}}
\subfigure{\includegraphics[angle=0,width=0.35\textwidth]{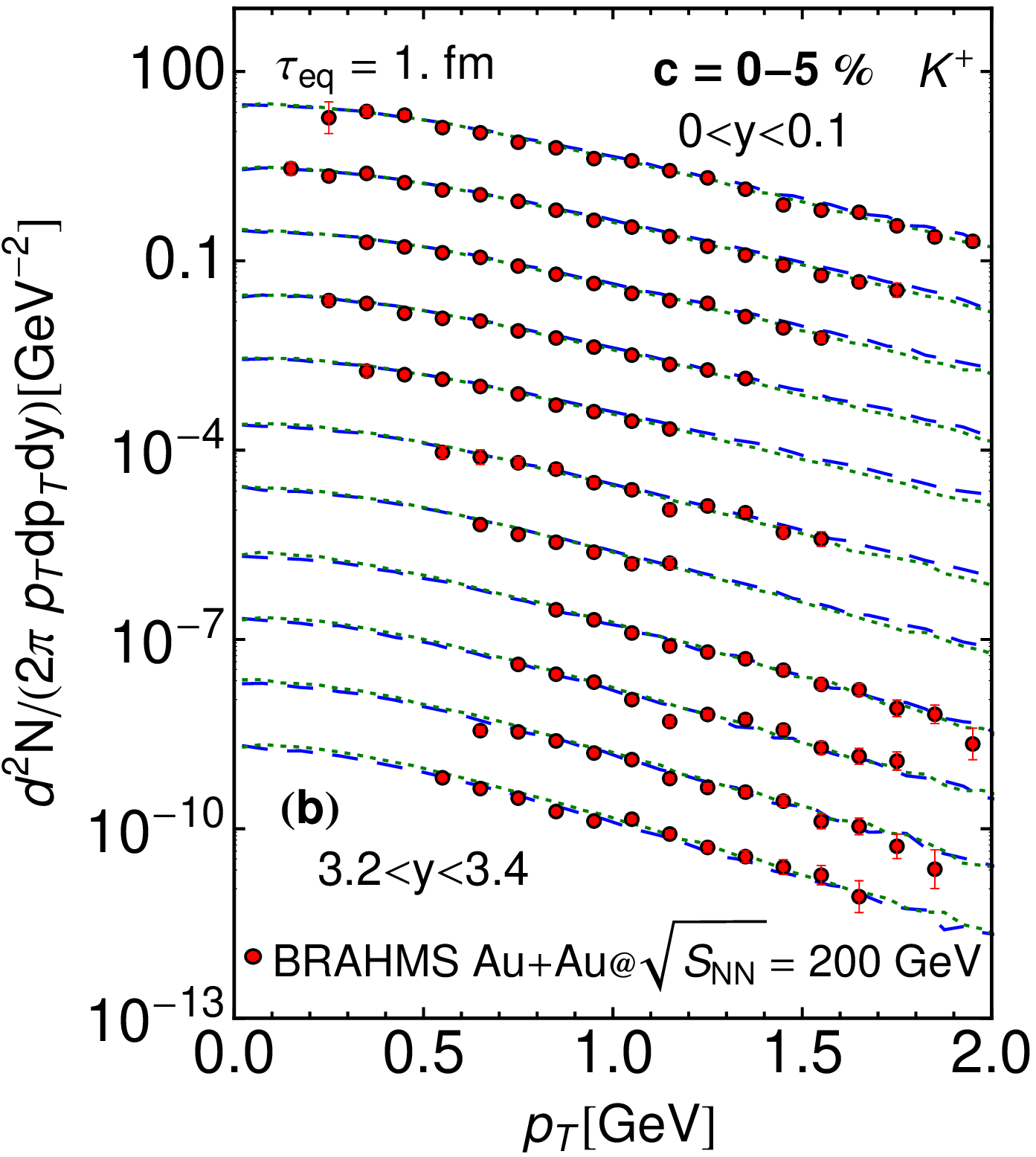}}
\end{center}
\caption{\small (Color online) Transverse momentum spectra of $\pi^{+}$ \textbf{(a)} and $K^{+}$ \textbf{(b)} for the centrality class $c=0-5$\% ($b=2.26$ fm) and for different rapidity windows. The model results are obtained with the time-scale parameter $\tau_{\rm eq}= 1.0$ fm. The results are obtained for two values of the initial anisotropy parameter: $x_{\rm 0}=100$ (dashed blue lines) and $x_{\rm 0}=0.032$ (dotted green lines). The model results are compared to the experimental data from BRAHMS \cite{Bearden:2004yx}. The spectra for different rapidity windows are successively rescaled down by factor $0.1$. 
}
\label{fig:rapptdistr_RHIC3}
\end{figure}

\subsection{Particle spectra}
\label{sect:partspec}

\subsubsection{Pseudorapidity distribution}
\label{sect:etadistr}

We start our discussion with the analysis of momentum distributions of charged and identified particles. These distributions include the feeding from unstable resonances whose  decays are implemented in \texttt{THERMINATOR}~\cite{Kisiel:2005hn,Chojnacki:2011hb}. 
At first, we analyze the pseudorapidity distributions of charged particles. The pseudorapidity is defined by the formula
\begin{eqnarray}
\eta_{PS} &=& \frac{1}{2} \ln \frac{p + p_{\parallel}}{p - p_{\parallel}}, \label{pseudoeta}
\end{eqnarray}
where $p$ is the momentum of a particle. 

In Fig.~\ref{fig:etadistr_RHIC} we show $dN_{\rm ch}/d\eta_{PS}$ for three different values of the time-scale parameter: $\tau_{\rm eq}=0.25$ fm {\bf (a)}, $\tau_{\rm eq}=0.5$ fm {\bf (b)}, and $\tau_{\rm eq}=1.0$ fm {\bf (c)}, and three values of the initial momentum anisotropy: $x_{\rm 0}=100$ (dashed blue lines), $x_{\rm 0}=1.0$ (solid black lines), and $x_{\rm 0}=0.032$ (dotted green lines). We note that in the case $x_0=1.0$ only the shortest value of the time-scale parameter $\tau_{\rm eq}=0.25$ fm is used --- this choice of the parameters leads to practically perfect-fluid behavior and may be treated as the reference calculation.

For $\tau_{\rm eq}=0.25$ fm, the results obtained with $x_{\rm 0}=100$ and $x_{\rm 0}=0.032$ are practically the same as those obtained with $x_{\rm 0}=1.0$, see Fig.~\ref{fig:etadistr_RHIC}~{\bf (a)}. As the duration of the off-equilibrium stage increases (larger values of $\tau_{\rm eq}$), the difference of pressures lasts longer, and the difference between the longitudinal and transverse expansion becomes larger. In particular, in the case $x_{\rm 0}=100$ the pseudorapidity distribution becomes a bit steeper in the region of forward and backward rapidities due to reduced longitudinal pressure, but the width of the central plateau is nearly unchanged, see Fig.~\ref{fig:etadistr_RHIC}~{\bf (b)}~and~{\bf (c)} . 

In Fig.~\ref{fig:etadistr_RHIC_m} we present the reference case with $x_{\rm 0}=1.0$ and $\tau_{\rm eq}=0.25$ fm together with the case $x_{\rm 0}=100$ and $\tau_{\rm eq}=1.0$ fm. In the latter case the initial energy density was {\it reduced} to $\varepsilon_{\rm i} = 41.8$  $\mathrm{GeV/fm^3}$ (from \mbox{$48.8$  $\mathrm{GeV/fm^3}$}) and the width of the plateau of the initial profile $f(\eta)$ was {\it increased} to $\Delta\eta = 1.5$ (from $\Delta\eta = 1.0$). In this way we obtain the same total multiplicity. We clearly see the improvement in the description of data compared to the case $x_{\rm 0}=100$ and $\tau_{\rm eq}=1.0$ fm shown in Fig.~\ref{fig:etadistr_RHIC} \textbf{(c)}.

Thus, the shape of pseudorapidity distributions may be quite well described if the initial fireball is transversally thermalized (i.e., the initial transverse pressure is much larger than the longitudinal pressure) but the initial profile parameters should be properly readjusted. It should be stressed, however, that the widths and forms of the {\it initial} longitudinal profiles (\ref{longitprof}) are poorly  known. Therefore, no definite conclusions about the length of the pre-equilibrium phase may be drawn from the correct description of the shape of the rapidity distributions alone, as it has been already noticed in  \cite{Bozek:2007qt}.

\begin{figure}[t]
\begin{center}
\subfigure{\includegraphics[angle=0,width=0.35\textwidth]{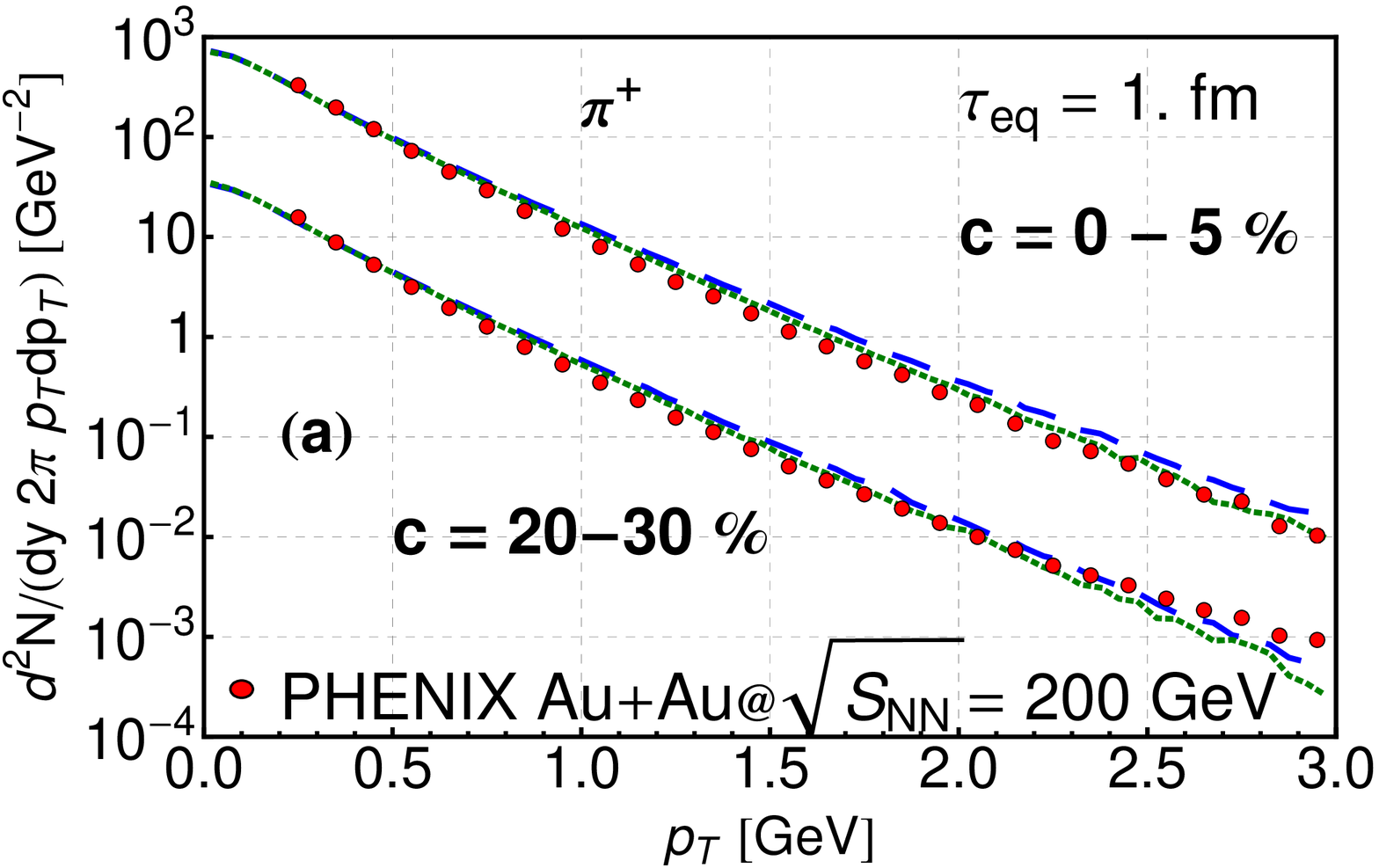}}
\subfigure{\includegraphics[angle=0,width=0.35\textwidth]{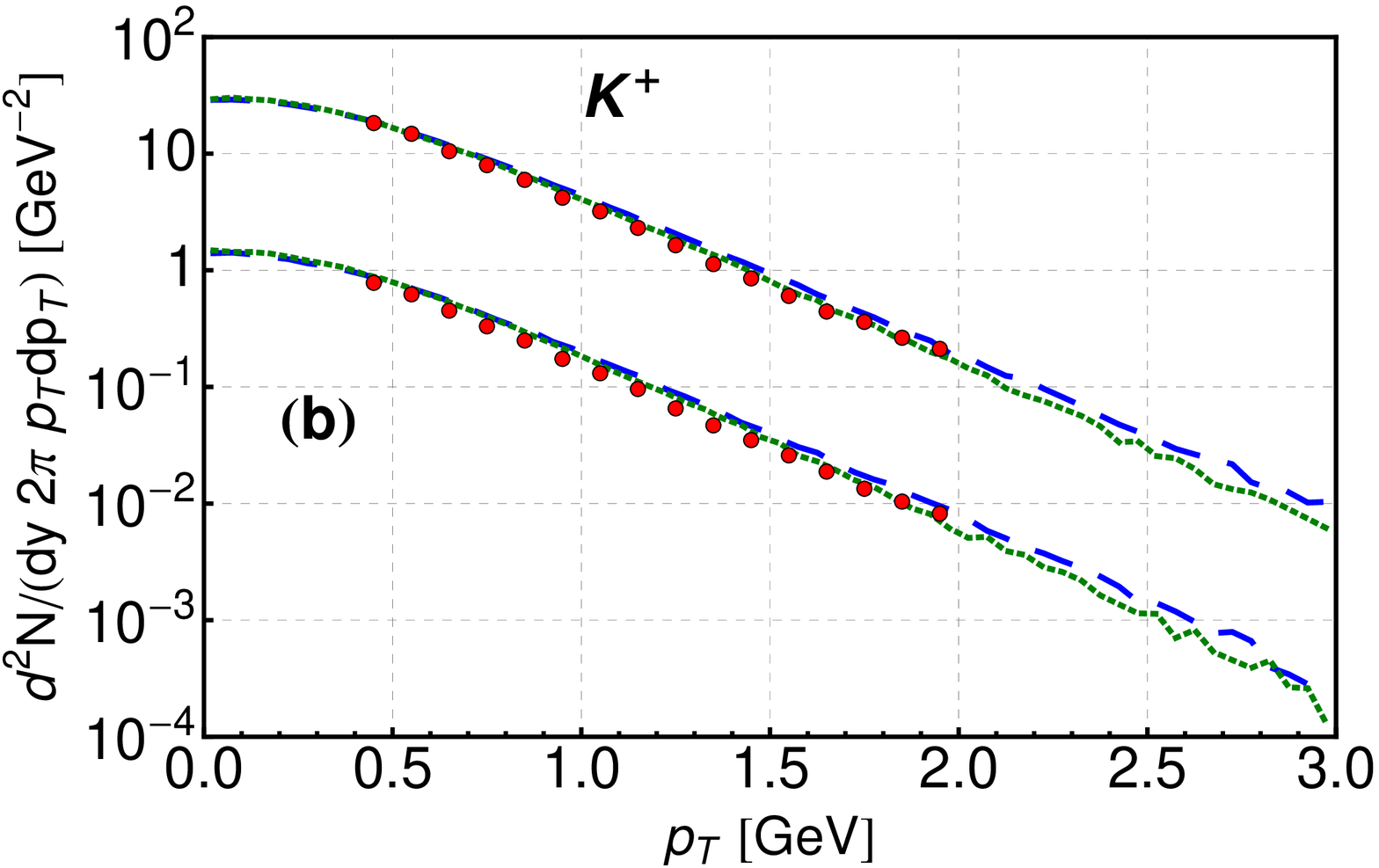}}
\subfigure{\includegraphics[angle=0,width=0.35\textwidth]{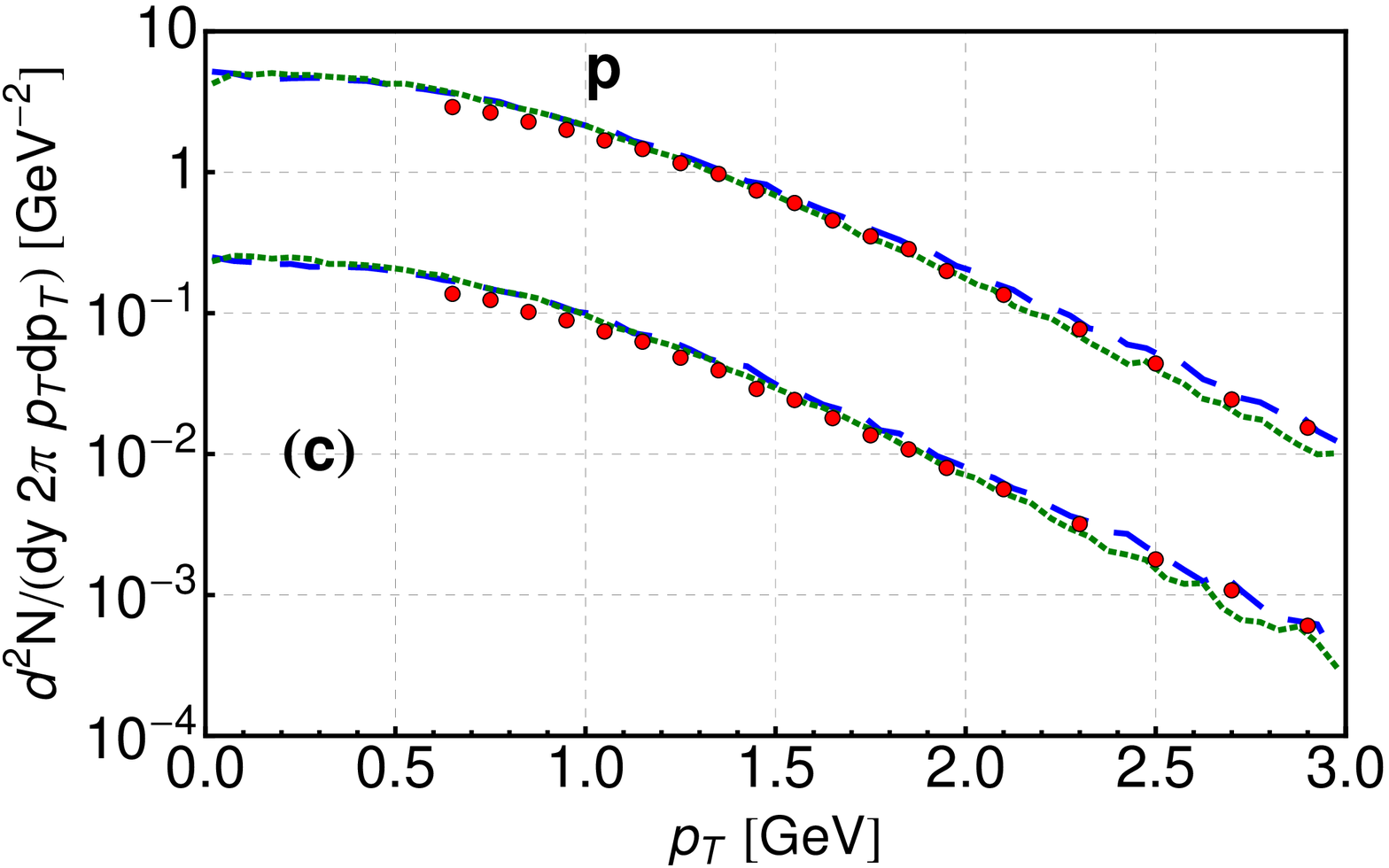}}
\end{center}
\caption{\small (Color online) Transverse momentum spectra of $\pi^{+}$ \textbf{(a)}, $K^{+}$ \textbf{(b)} and $p$ \textbf{(c)} for the two centralities: $c=0-5$\% ($b=2.26$ fm) and $c=20-30$\% ($b=7.16$ fm), compared to the experimental data from PHENIX \cite{Adler:2003cb}. The spectra for $c=20-30$\% are scaled down by factor $0.1$. The results for $x_0=100$ (dashed blue lines) and $x_0=0.032$ (dotted green lines) are obtained with the time-scale parameter $\tau_{\rm eq}= 1.0$ fm.
}
\label{fig:centrptdistr1_RHIC3}
\end{figure}

\begin{figure}[t]
\begin{center}
\subfigure{\includegraphics[angle=0,width=0.35\textwidth]{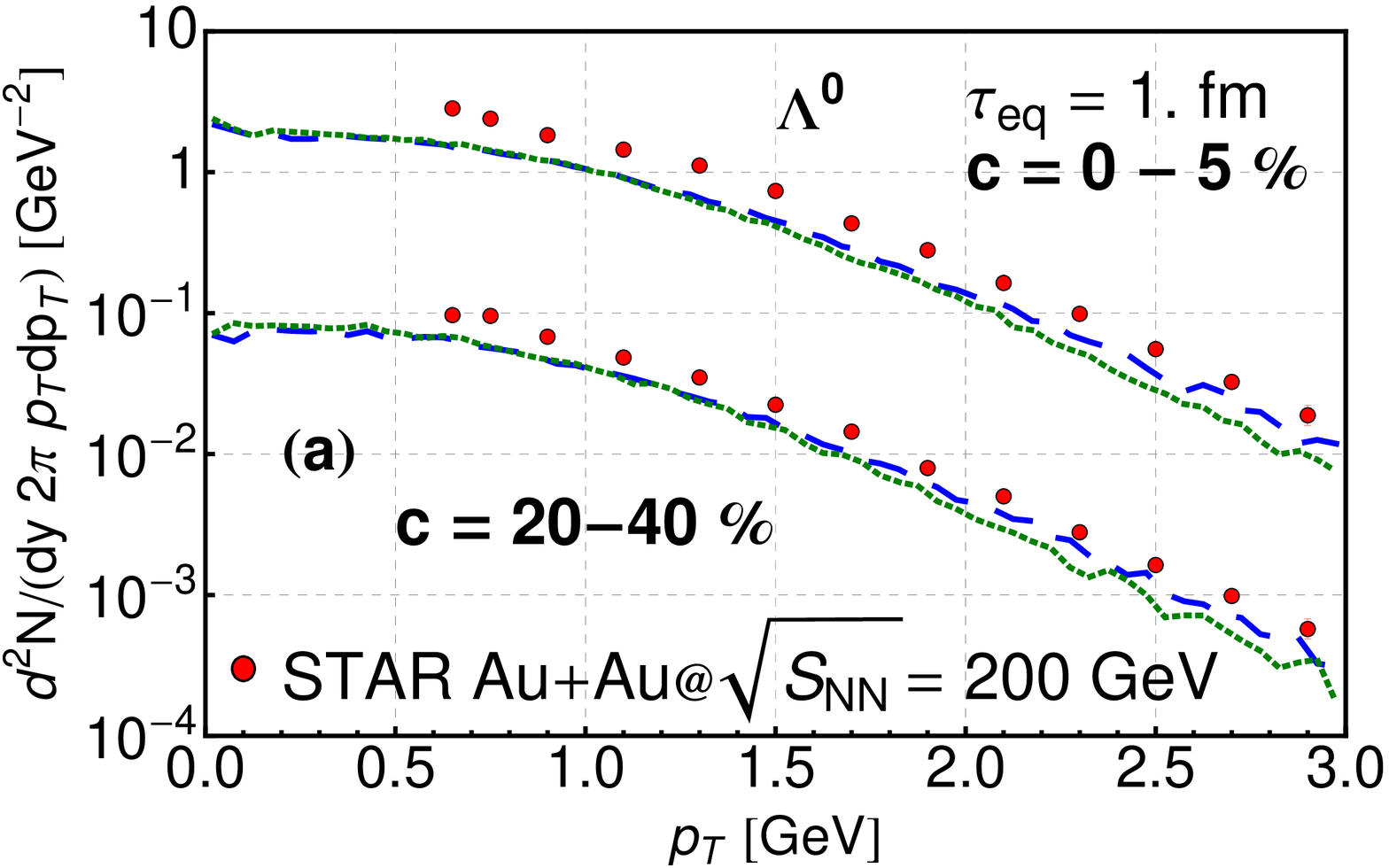}}
\subfigure{\includegraphics[angle=0,width=0.35\textwidth]{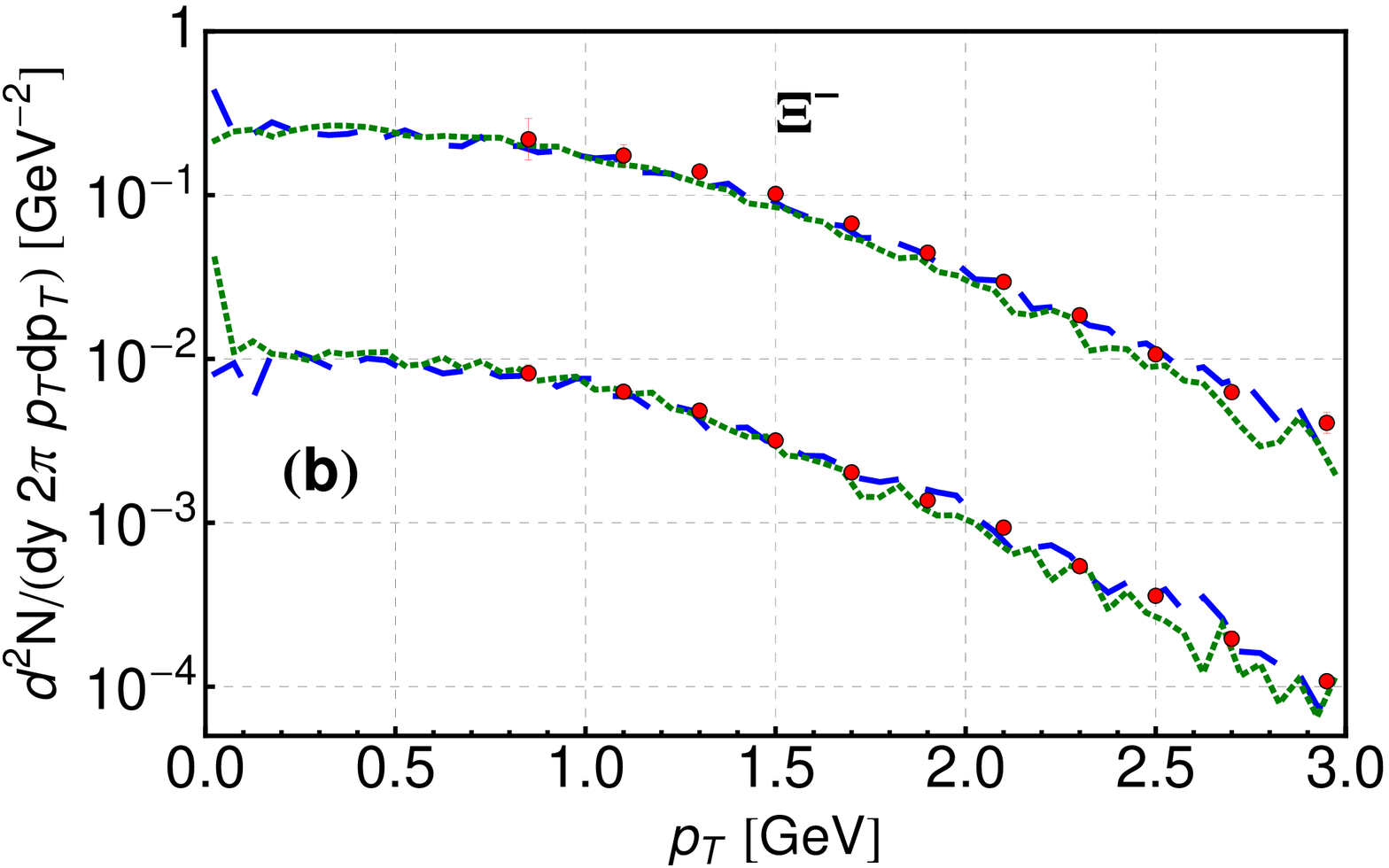}}
\subfigure{\includegraphics[angle=0,width=0.35\textwidth]{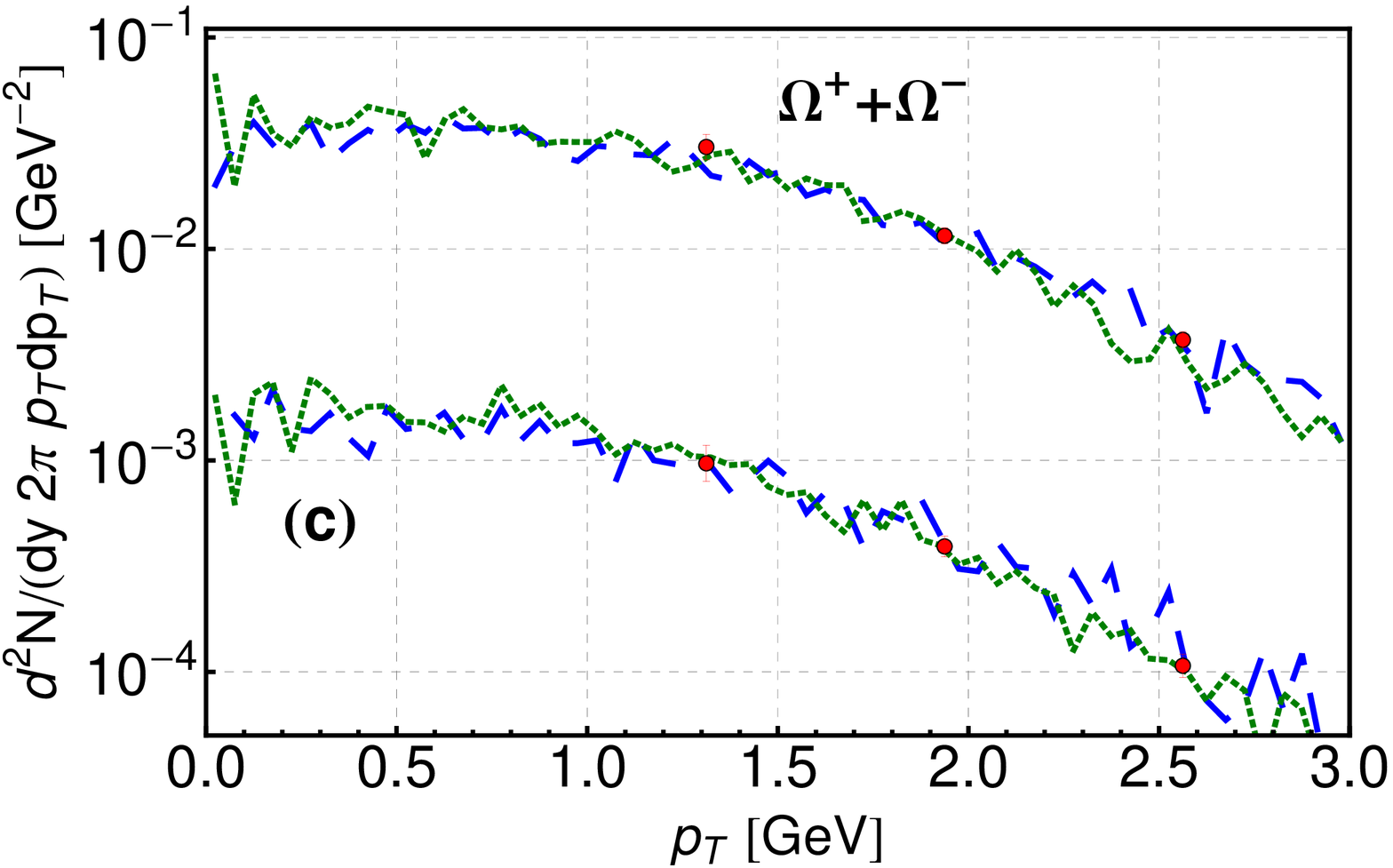}}
\end{center}
\caption{\small (Color online) Transverse-momentum spectra of $\Lambda^{0}$  \textbf{(a)}, $\Xi^{-}$  \textbf{(b)} and $\Omega^{-}+\Omega^{+}$ \textbf{(c)} for the two centralities: \mbox{$c=0-5$\%} and $c=20-40$\% ($b=7.84$ fm), compared to the experimental data from STAR \cite{Adams:2006ke}. The spectra for $c=20-40$\% are scaled down by factor $0.1$. The results for $x_0=100$ (dashed blue lines) and $x_0=0.032$ (dotted green lines) are obtained with the time-scale parameter $\tau_{\rm eq}= 1.0$ fm. The $\Lambda$ feed-down corrections for proton spectra have been applied. The $\Lambda$ spectra were feed-down corrected for $\Xi$ and $\Omega$.
}
\label{fig:centrptdistr2_RHIC3}
\end{figure}

\subsubsection{Transverse-momentum spectra}
\label{sect:transmomsp}

In Fig.~\ref{fig:rapptdistr_RHIC3} we present the transverse-momentum spectra of positive pions \textbf{(a)} and kaons \textbf{(b)} for the centrality class $c=0-5$\% ($b=2.26$ fm) and for different rapidity windows. The model calculations are done with $\tau_{\rm eq}= 1.0$ fm. The model results are compared to the BRAHMS data \cite{Bearden:2004yx}. We reproduce well the data in the two different cases: $x_{\rm 0}=100$ (dashed blue lines) and $x_{\rm 0}=0.032$ (dotted green lines). The correct description of the  $p_{T}$ spectra in a broad range of rapidities is usually understood as an indication for a high level of thermalization of the entire fireball \cite{Bozek:2009ty}. Our calculation shows that after appropriate rescaling of the initial central energy density, the spectra are almost insensitive to initial values of anisotropy and to the studied length of the anisotropic stage (note a relatively large value of $\tau_{\rm eq}$ in Fig.~\ref{fig:rapptdistr_RHIC3}). The small discrepancies in normalizations between different rapidity windows are mainly due to discrepancies between theoretical and experimental $dN_{\rm ch}/d\eta_{PS}$ profiles, see Fig.~\ref{fig:etadistr_RHIC}~\textbf{(c)}.

In Fig.~\ref{fig:centrptdistr1_RHIC3} we present the transverse-momentum spectra of positive pions \textbf{(a)}, kaons \textbf{(b)}, and protons \textbf{(c)} at midrapidity ($|\mathrm{y}| < 1$) for the time-scale parameter $\tau_{\rm eq}= 1.0$ fm and for the two centralities: $c=0-5$\% \mbox{($b=2.26$ fm)} and $c=20-30$\% ($b=7.16$ fm). The model results are compared to the data from PHENIX \cite{Adler:2003cb}. We can see that the model reproduces very well the slope of the spectra of $K^{+}$'s and protons up to $p_{T}=3$ GeV. The model pion spectra for \mbox{$c=20-30$\%} are underestimated for $p_{T}>2.5$ GeV. This is an expected feature, since hard particles require more rescatterings to thermalize and this region is out of applicability of hydrodynamics. 

The correct normalization of spectra for different hadron species has been slightly improved by introducing chemical potentials at freeze-out, which were obtained in the framework of a thermal model using a complete treatment of resonances \cite{Florkowski:2001fp,Baran:2003nm}. The reproduction of the convex shape of the pion spectra at low $p_{T}$ is a consequence of feeding from resonance decays which are implemented in \texttt{THEMINATOR} \cite{Kisiel:2005hn,Chojnacki:2011hb}. 

Figure~\ref{fig:centrptdistr2_RHIC3} presents the transverse-momentum spectra of hyperons for the centralities $c=0-5$\% \mbox{($b=2.26$ fm)} and $c=20-40$\% \mbox{($b=7.84$ fm)} compared to the data from STAR \cite{Adams:2006ke}. Although our freeze-out temperature is a bit lower than that used in thermal models, the slopes of the hyperon spectra are reproduced very well. The normalization of $\Lambda$'s is too small, the effect that may indicate their higher freeze-out temperature. Figures~\ref{fig:centrptdistr1_RHIC3} and \ref{fig:centrptdistr2_RHIC3} show again a small sensitivity of the model predictions to initial anisotropy, which is consistent with conclusions based on Fig.~\ref{fig:rapptdistr_RHIC3}. 

One would naively expect that the increase (decrease) of the pressure ratio $P_{\perp}/P_{\parallel}$ in the case $x_{\rm 0}=100$ ($x_{\rm 0}=0.032$),  compared  to the perfect-fluid case,  should lead to slower (faster) cooling of the system in the transverse direction. However the renormalization of the initial energy density (according to Table~\ref{table:multiplicity}) in order to obtain the same final multiplicity leads to a reduction (amplification) of the transverse flow produced during expansion in the anisotropic phase \cite{Bozek:2010aj}. This explicitly confirms the phenomenon of universality of the flow,  predicting that the overall growth of the flow is the same regardless of the pressure anisotropy of the energy-momentum tensor \cite{Vredevoogd:2008id}.

\subsection{Directed and elliptic flows}
\label{sect:anisoflow}
%

\begin{figure}[t]
\begin{center}
\subfigure{\includegraphics[angle=0,width=0.4\textwidth]{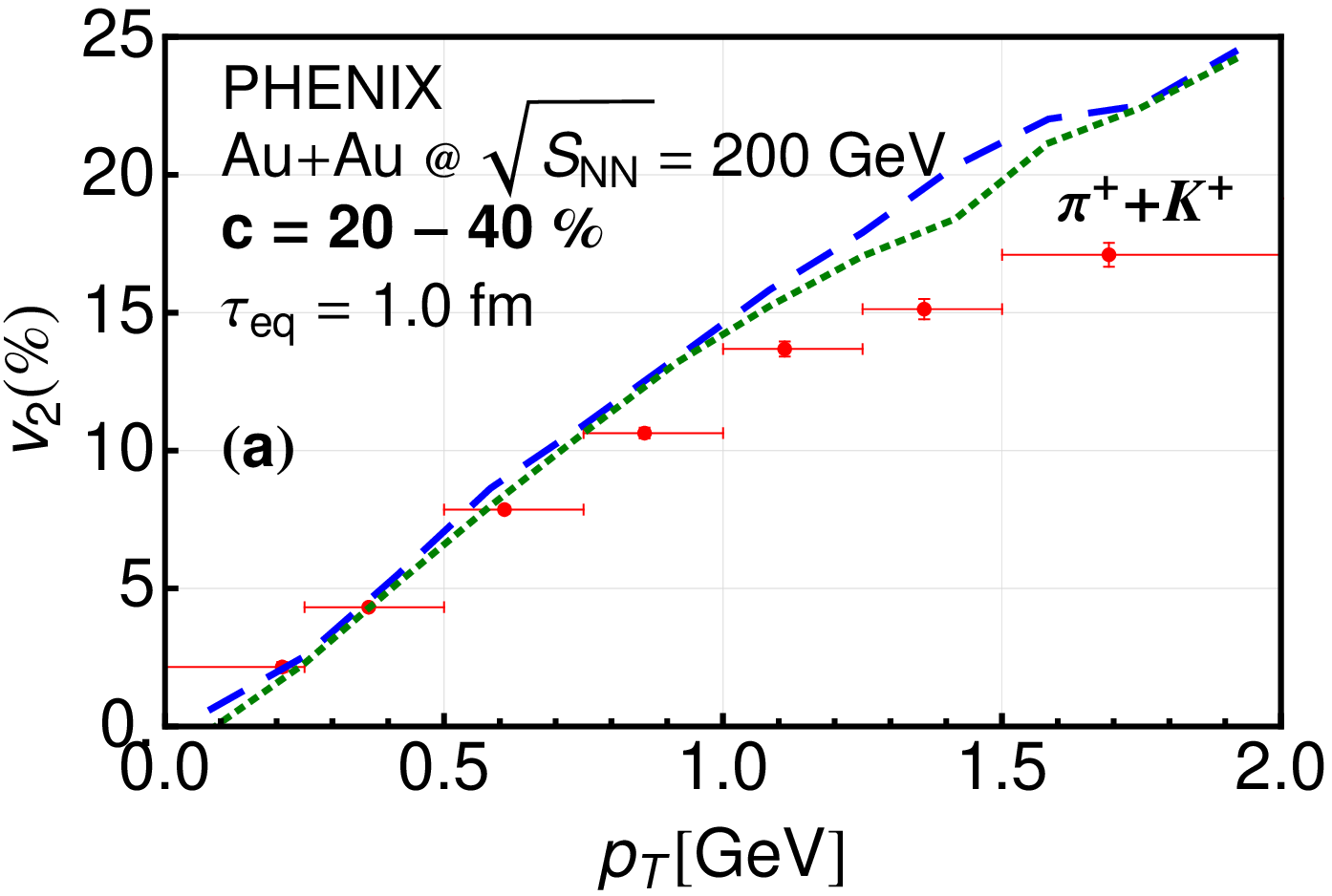}}
\subfigure{\includegraphics[angle=0,width=0.4\textwidth]{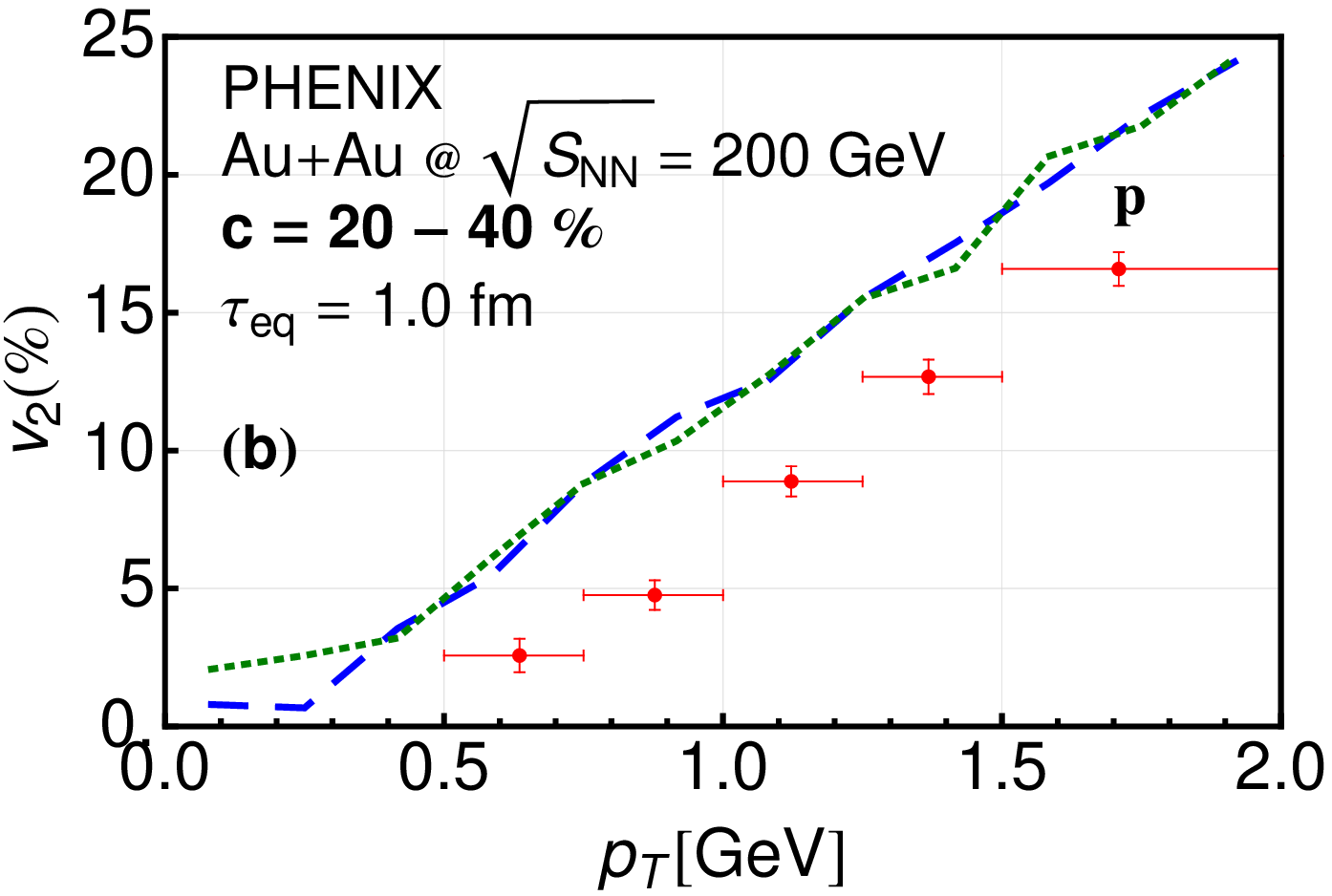}}
\end{center}
\caption{\small (Color online) Transverse-momentum dependence of the elliptic flow coefficient $v_2$ of $\pi^{+}+K^{+}$ \textbf{(a)} and protons \textbf{(b)} calculated for the centrality $c=20-40$\% ($b=7.84$ fm) at midrapidity for the time-scale parameter $\tau_{\rm eq}= 1.0$ fm, and for the two values of the initial anisotropy parameter: $x_0=100$ (dashed blue lines) and $x_0=0.032$ (dotted green lines). The results are compared to the PHENIX Collaboration data (red dots) \cite{Adler:2003kt}. The presented errors are statistical. Horizontal bars denote the $p_T$ bins.
}
\label{fig:ellflowpT}
\end{figure}
\begin{figure}[t]
\begin{center}
\subfigure{\includegraphics[angle=0,width=0.35\textwidth]{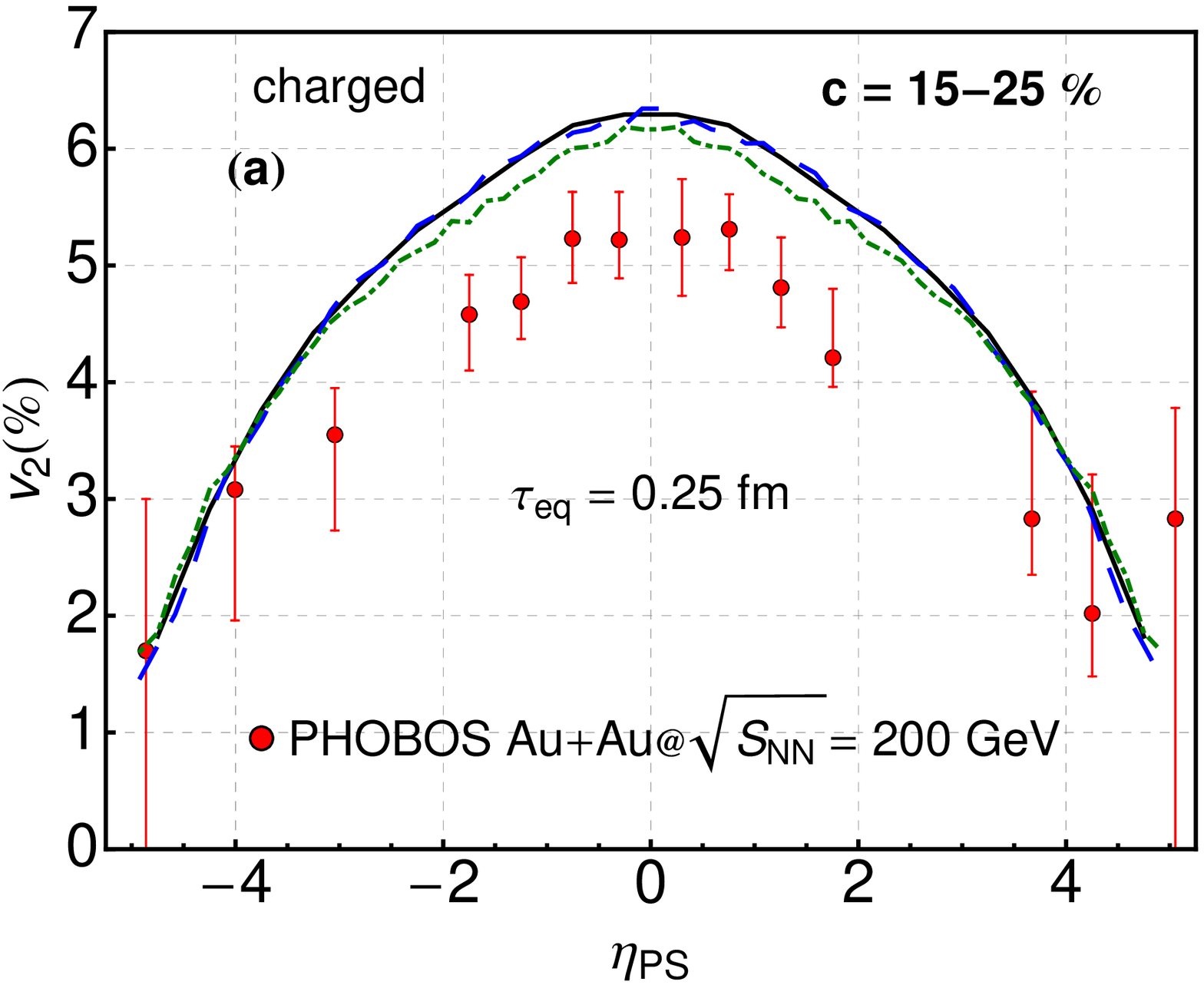}}
\subfigure{\includegraphics[angle=0,width=0.35\textwidth]{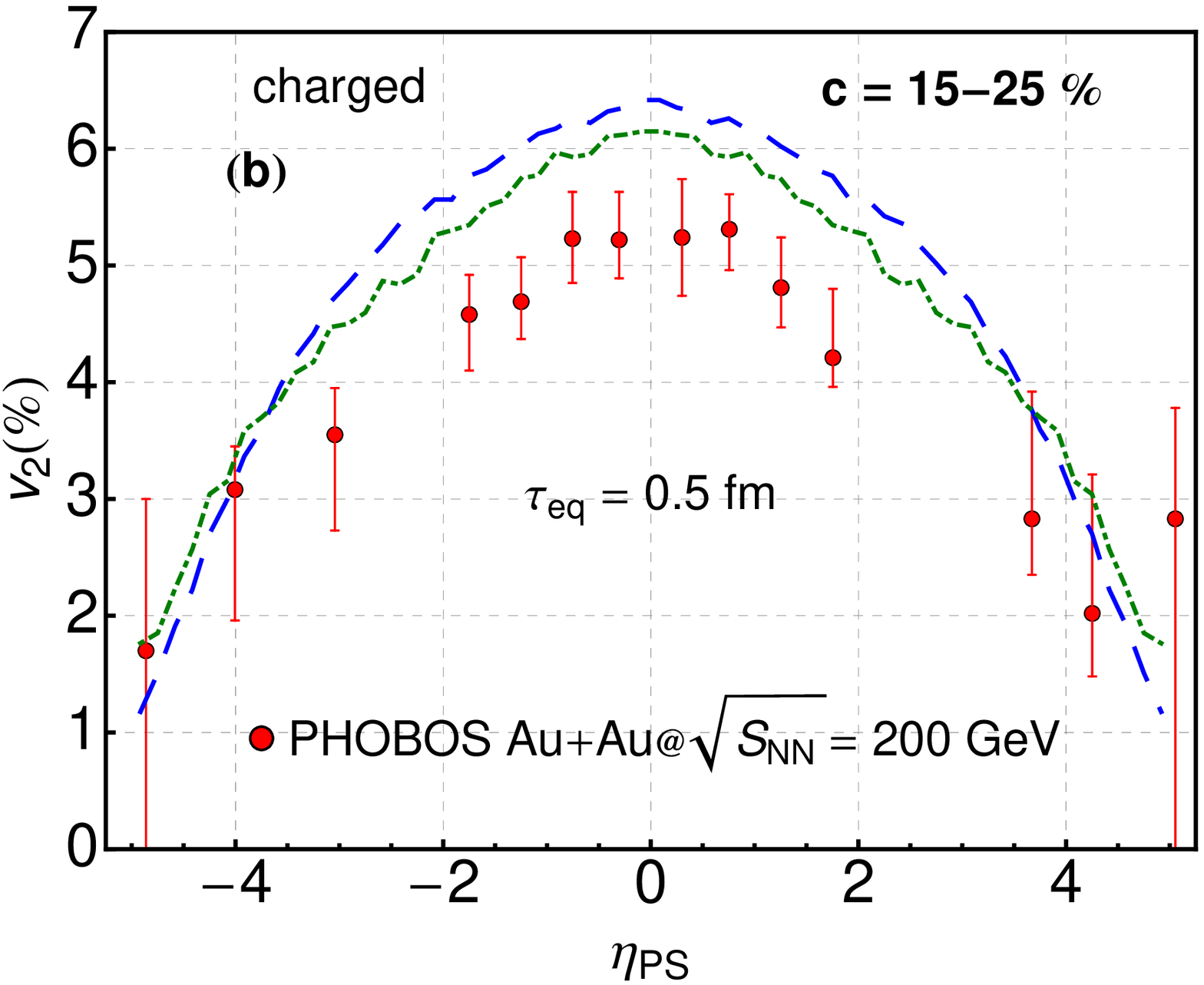}}
\subfigure{\includegraphics[angle=0,width=0.35\textwidth]{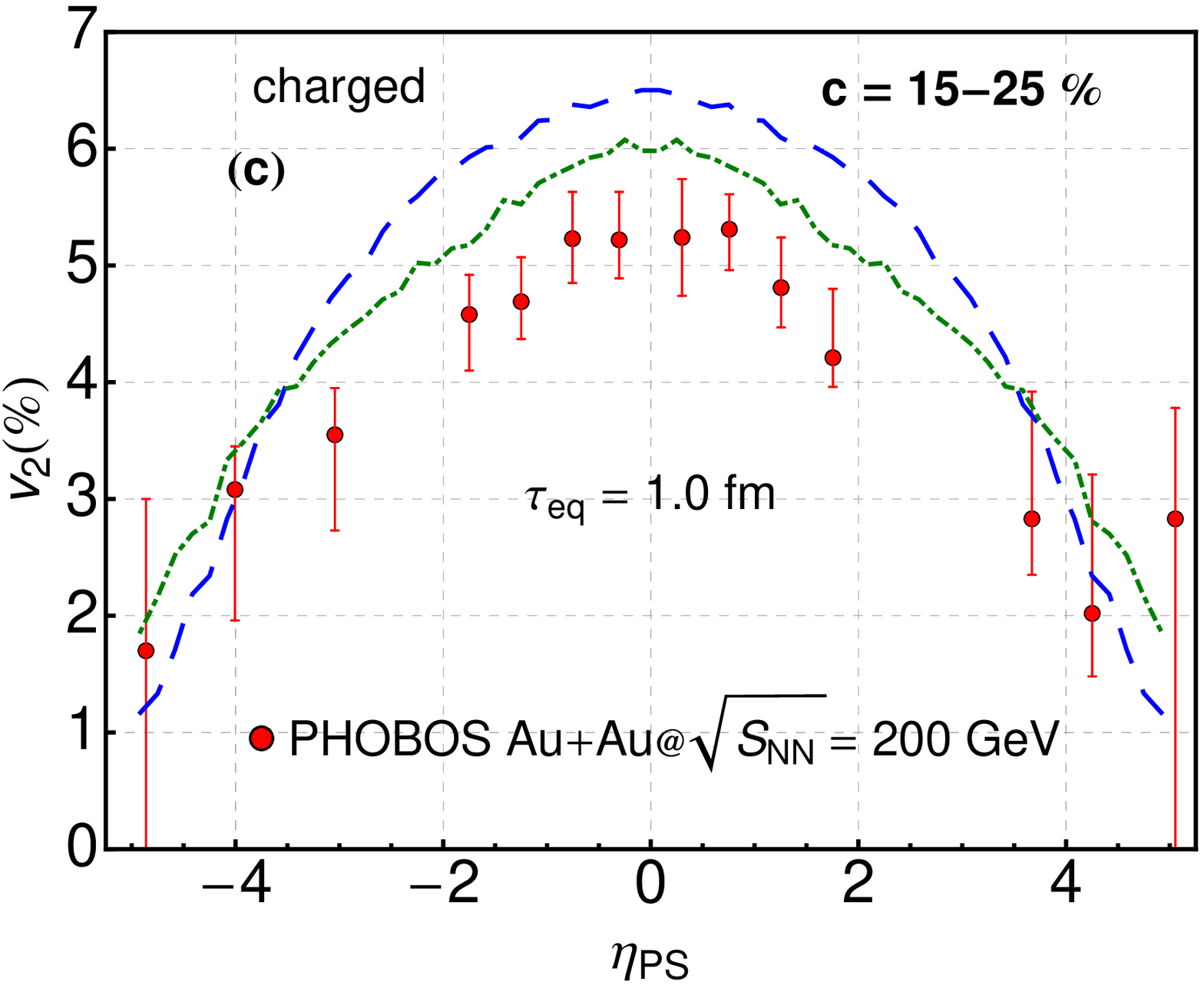}}
\end{center}
\caption{\small (Color online) Pseudorapidity dependence of the $p_{T}$-integrated elliptic flow of charged particles calculated for the centrality $c=15-25$\%, three time-scale parameters: $\tau_{\rm eq}=0.25$ fm \textbf{(a)}, $\tau_{\rm eq}=0.5$ fm \textbf{(b)}, $\tau_{\rm eq}=1.0$ fm \textbf{(c)}, and three values of the initial anisotropy parameter: $x_0=100$ (dashed blue lines), $x_0=1.0$ (solid black lines), and $x_0=0.032$ (dotted green lines). The results are compared to the PHOBOS Collaboration data (red dots) \cite{Back:2004mh}. }
\label{fig:ellfloweta}
\end{figure}

The significant amount of anisotropic flows is built at the very early stages of the evolution of matter, when the gradients of pressures are the largest due to the large {\it eccentricity} of the source. The inclusion of early non-equilibrium stages of the collisions may significantly change the momentum anisotropy observed in the final spectra, making $v_n$ coefficients sensitive probes for the possible existence of non-equilibrium stages. 
\begin{figure}[t]
\begin{center}
\subfigure{\includegraphics[angle=0,width=0.35\textwidth]{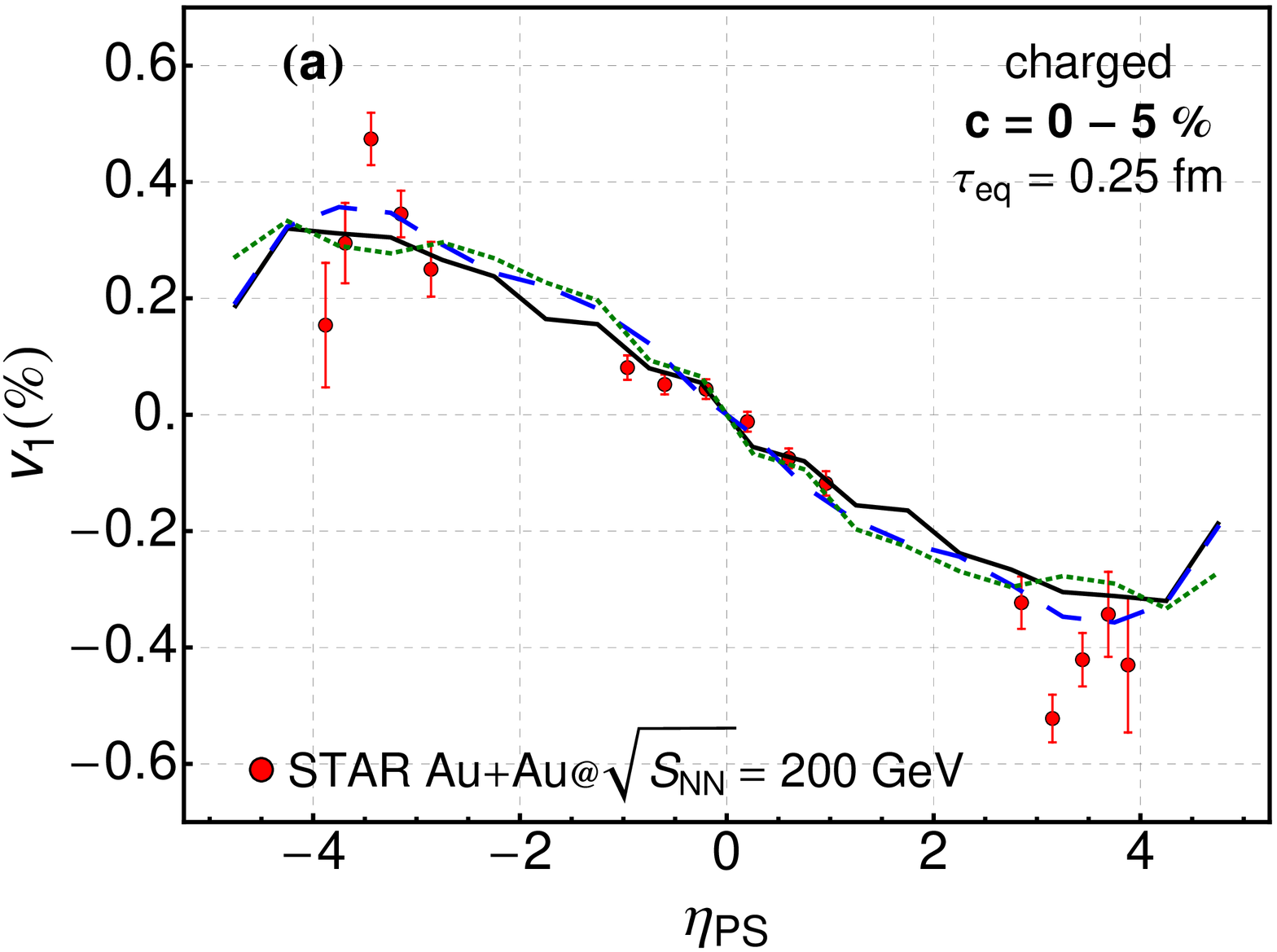}}
\subfigure{\includegraphics[angle=0,width=0.35\textwidth]{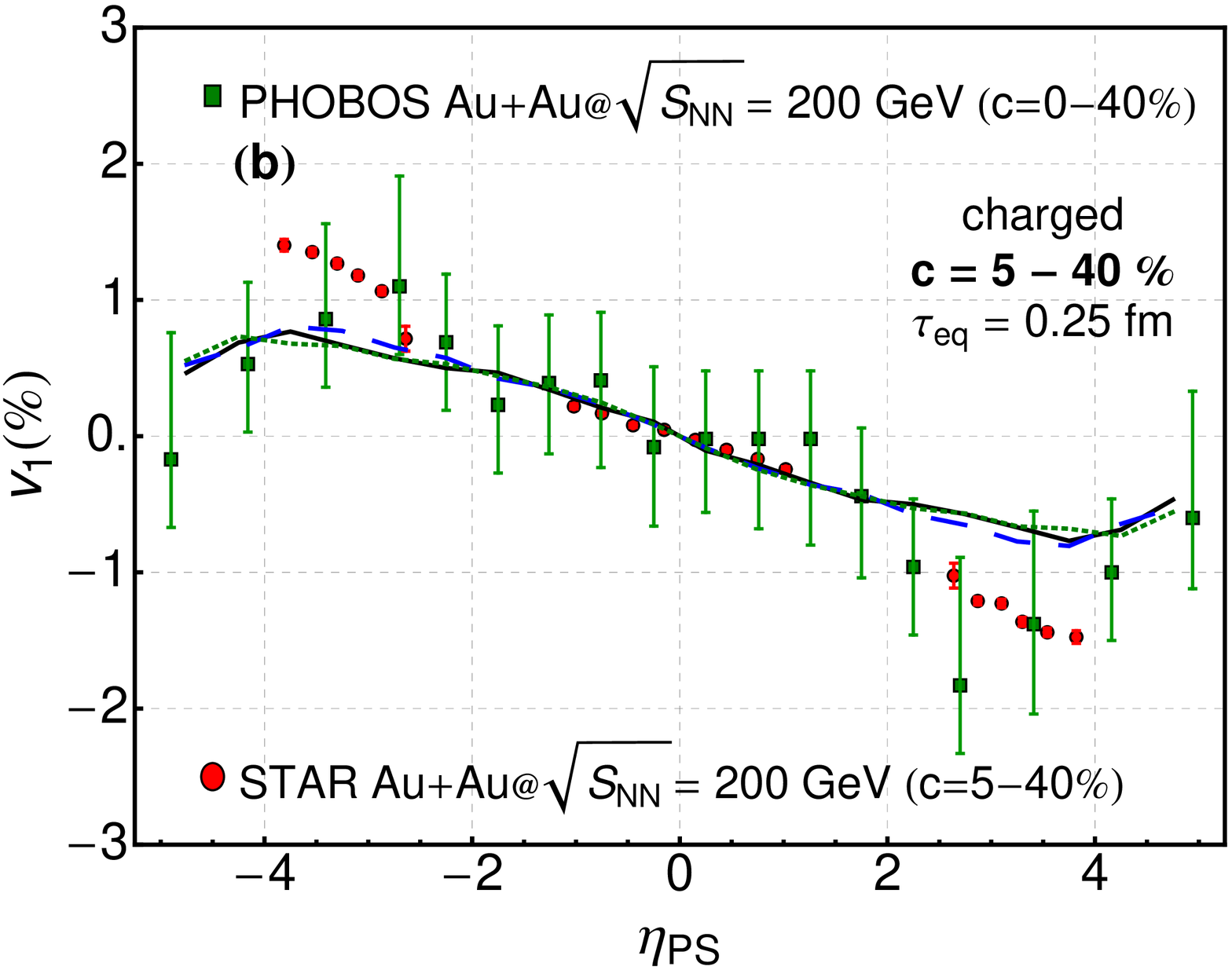}}
\end{center}
\caption{\small (Color online) Pseudorapidity dependence of the directed flow of charged particles for the two centrality bins:  $c=0-5$\% \textbf{(a)} and $c=5-40$\% \textbf{(b)}, the time-scale parameter $\tau_{\rm eq}=0.25$ fm, and for three values of the initial anisotropy parameter: $x_0=100$ (dashed blue lines), $x_0=1.0$ (solid black lines), and $x_0=0.032$ (dotted green lines). The results are compared to the experimental data from STAR (red dots) \cite{Abelev:2008jga} and PHOBOS (green squares) \cite{Back:2005pc}. 
}
\label{fig:dirfloweta11}
\end{figure}

\begin{figure}[t]
\begin{center}
\subfigure{\includegraphics[angle=0,width=0.35\textwidth]{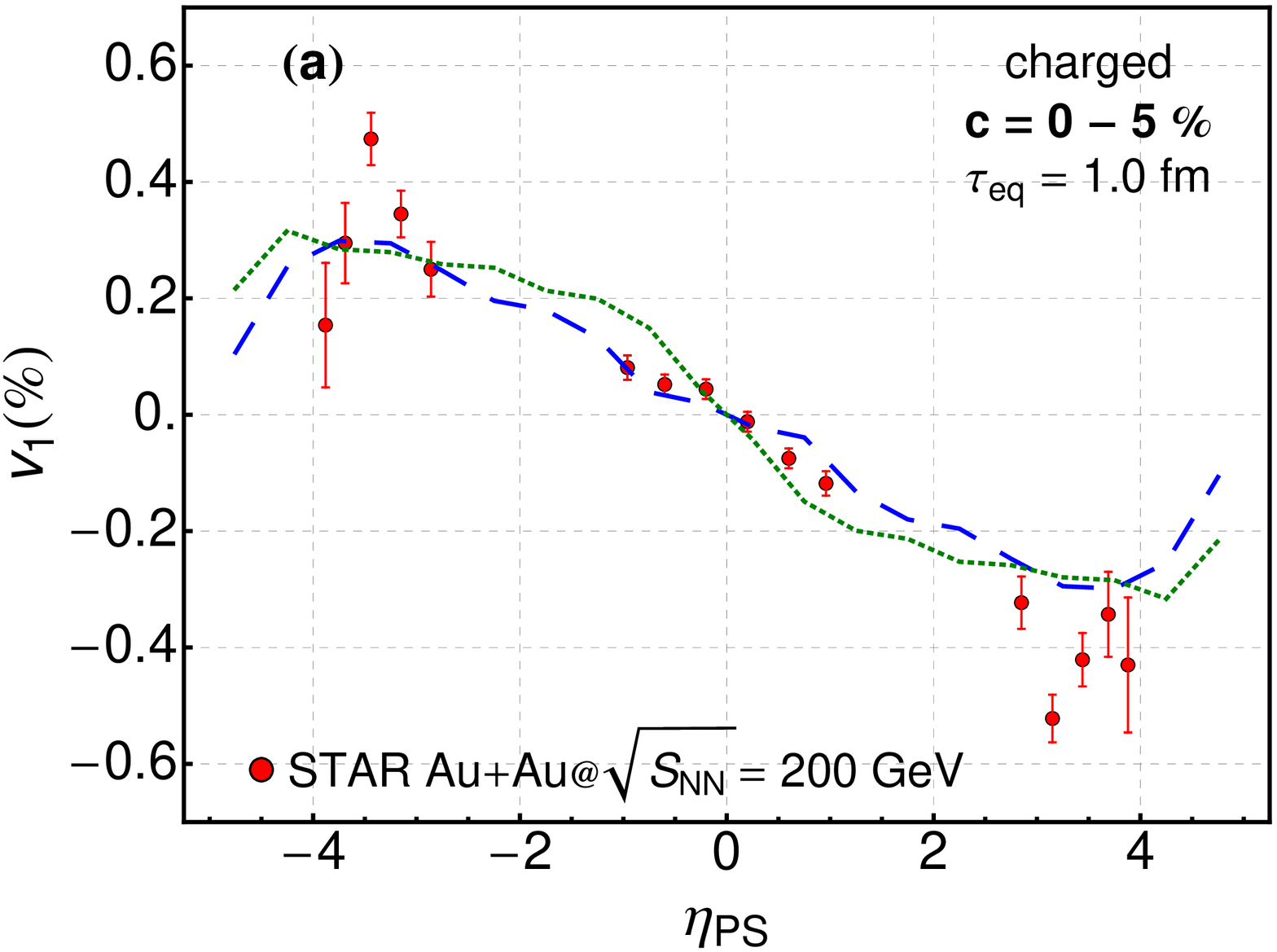}}
\subfigure{\includegraphics[angle=0,width=0.35\textwidth]{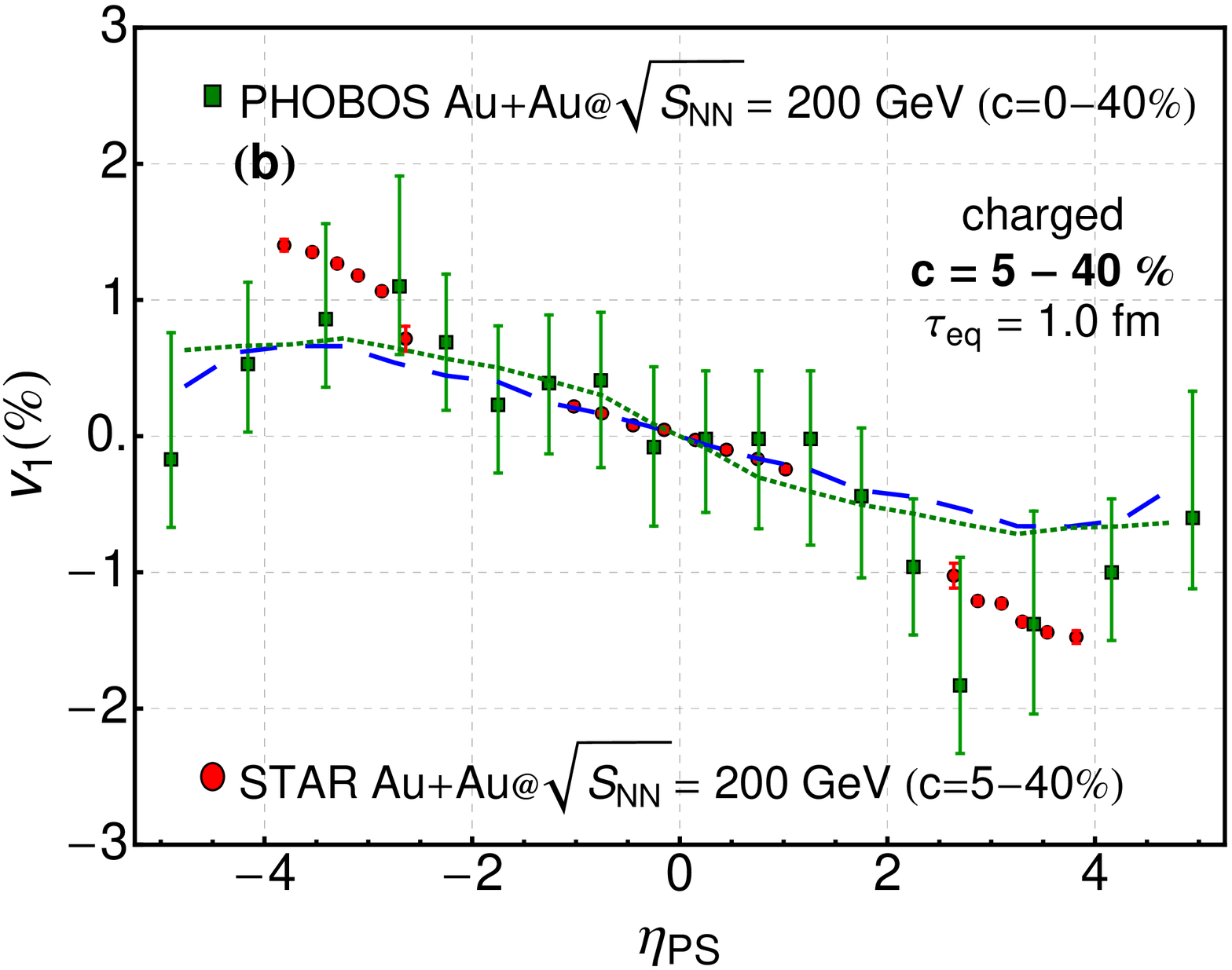}}
\end{center}
\caption{\small (Color online) The same as Fig.~\ref{fig:dirfloweta11} but for the time-scale parameter $\tau_{\rm eq}=1.0$ fm. 
}
\label{fig:dirfloweta12}
\end{figure}

\subsubsection{Elliptic flow}
\label{sect:3Dv2}

\par In the midrapidity region of non-central Au+Au collisions, the main information about the momentum anisotropy  of the produced particles is contained in the $v_2$ coefficient. In Fig.~\ref{fig:ellflowpT} \textbf{(a)} we show the elliptic flow of $\pi^{+}+K^{+}$ as a function of the transverse momentum $p_T$. The model results obtained for the centrality $c=20-40$\% ($b=7.84$ fm) and with the time-scale parameter $\tau_{\rm eq}= 1.0$ fm are compared to the data from PHENIX \cite{Adler:2003kt}. We observe that the model reproduces well the data up to $p_{T}=1$ GeV. The saturation for higher $p_{T}$ is not reproduced, since it requires inclusion of shear viscosity effects \cite{Bozek:2011ua}. Again, the model results weakly depend on the initial value of anisotropy, since the increase (decrease) of transverse pressure in the initial anisotropic stage is compensated by the decrease (increase) of the initial energy density. For completeness, in Fig.~\ref{fig:ellflowpT} \textbf{(b)} we show the $p_{T}$-dependence of the elliptic flow of protons. The mass-splitting of $v_2$ is not reproduced. The elliptic flow of protons may be better described by introducing bulk viscosity \cite{Bozek:2009dw}. 

\par In Fig.~\ref{fig:ellfloweta} we present the $p_{T}$-integrated elliptic flow as a function of pseudorapidity. The model calculations are done for the centrality class $c=15-25$\%, three time-scale parameters, and three values of the initial anisotropy parameter. The model results are compared to the PHOBOS data \cite{Back:2004mh}. We observe that the model overshoots the data by about 20\% in the reference case where $x_{\rm 0}=1.0$ and $\tau_{\rm eq}= 0.25$ fm (solid black line). The peaked shape visible in the data is not well reproduced. 

We find that the elliptic flow increases if the transverse pressure dominates. In this case, the elliptic flow is also steeper in pseudorapidity. It is easy to understand this behavior, since the reduction of longitudinal pressure leads to slower expansion of the fireball in the longitudinal direction. In the case of dominating longitudinal pressure the behavior is opposite. The strength of this effect depends on the length of the anisotropic stage. Thus, we conclude that the $p_{T}$-integrated $v_2$ is sensitive to the early anisotropic stages.

\begin{figure}[t]
\begin{center}
\subfigure{\includegraphics[angle=0,width=0.35\textwidth]{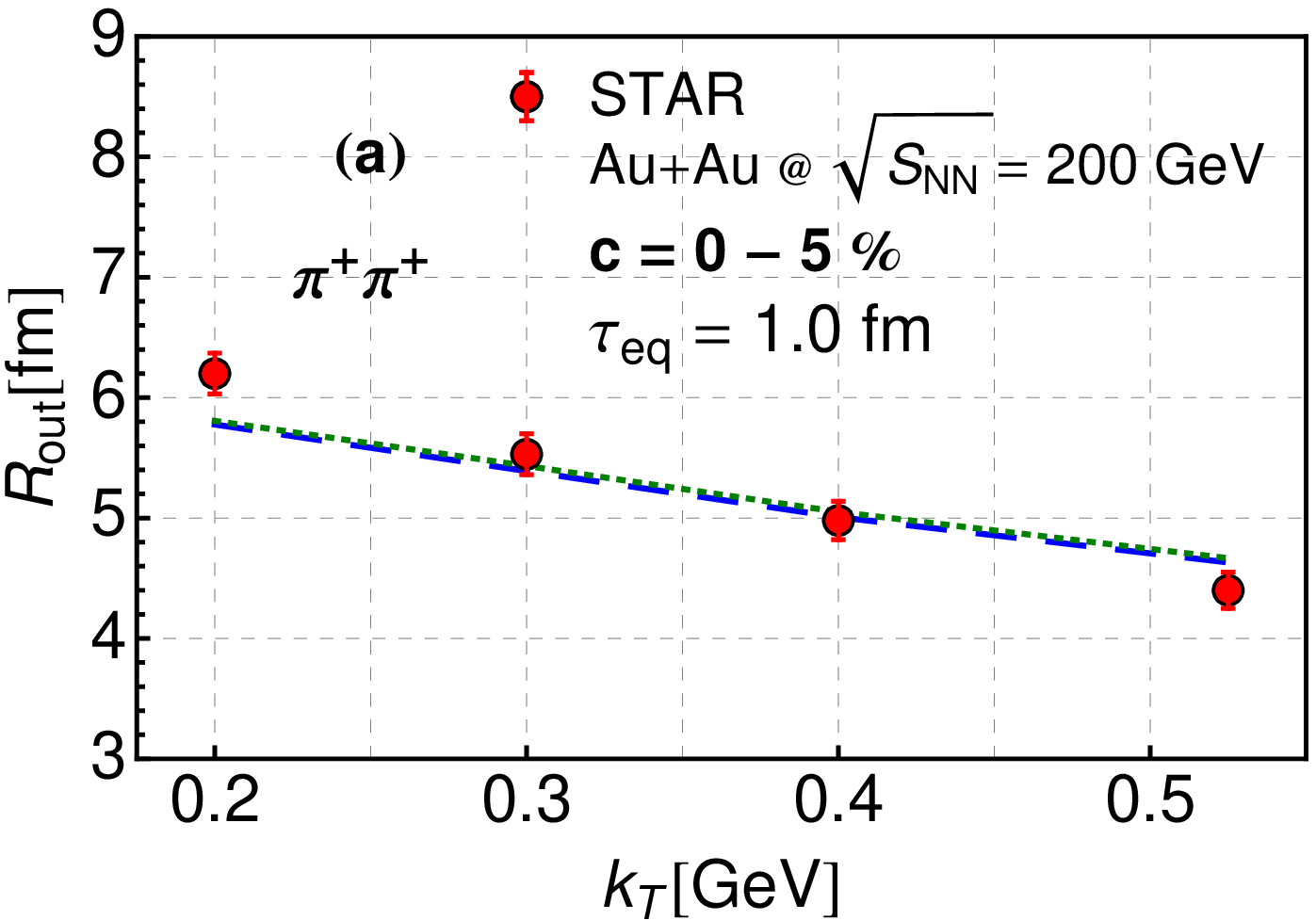}}
\subfigure{\includegraphics[angle=0,width=0.35\textwidth]{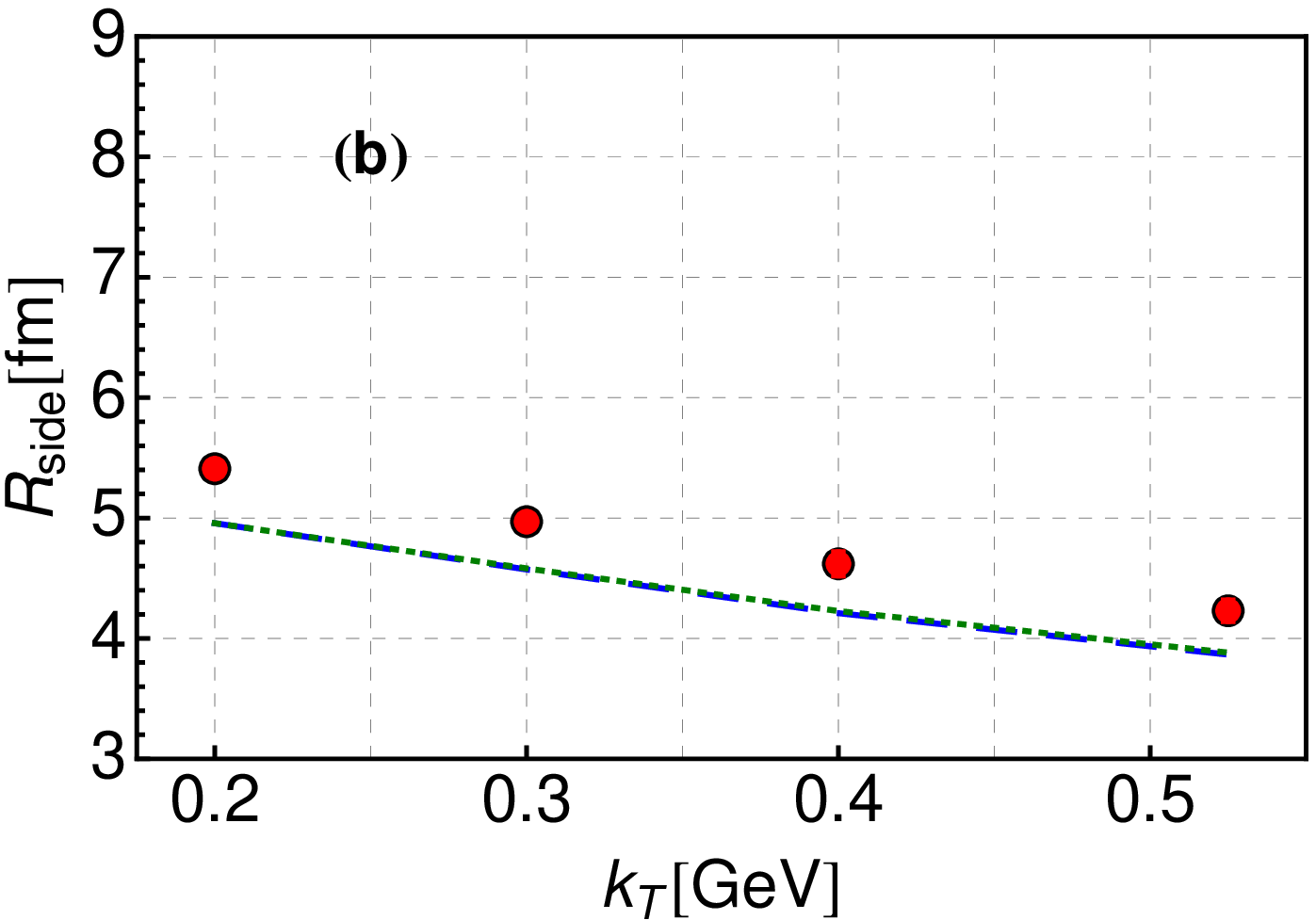}}
\end{center}
\caption{\small (Color online) The HBT correlation radii $R_{\rm out}$ \textbf{(a)} and $R_{\rm side}$ \textbf{(b)} of positive pions as functions of the total transverse momentum of the pair for the centrality $c=0-5$\%, the time-scale parameter $\tau_{\rm eq}=1.0$ fm, and for two values of the initial anisotropy: $x_0=100$ (dashed blue lines) and $x_0=0.032$ (dotted green lines). The results are compared to experimental data from STAR Collaboration (red dots) \cite{Adams:2004yc}. 
}
\label{fig:hbt11}
\end{figure}

\begin{figure}[t]
\begin{center}
\subfigure{\includegraphics[angle=0,width=0.35\textwidth]{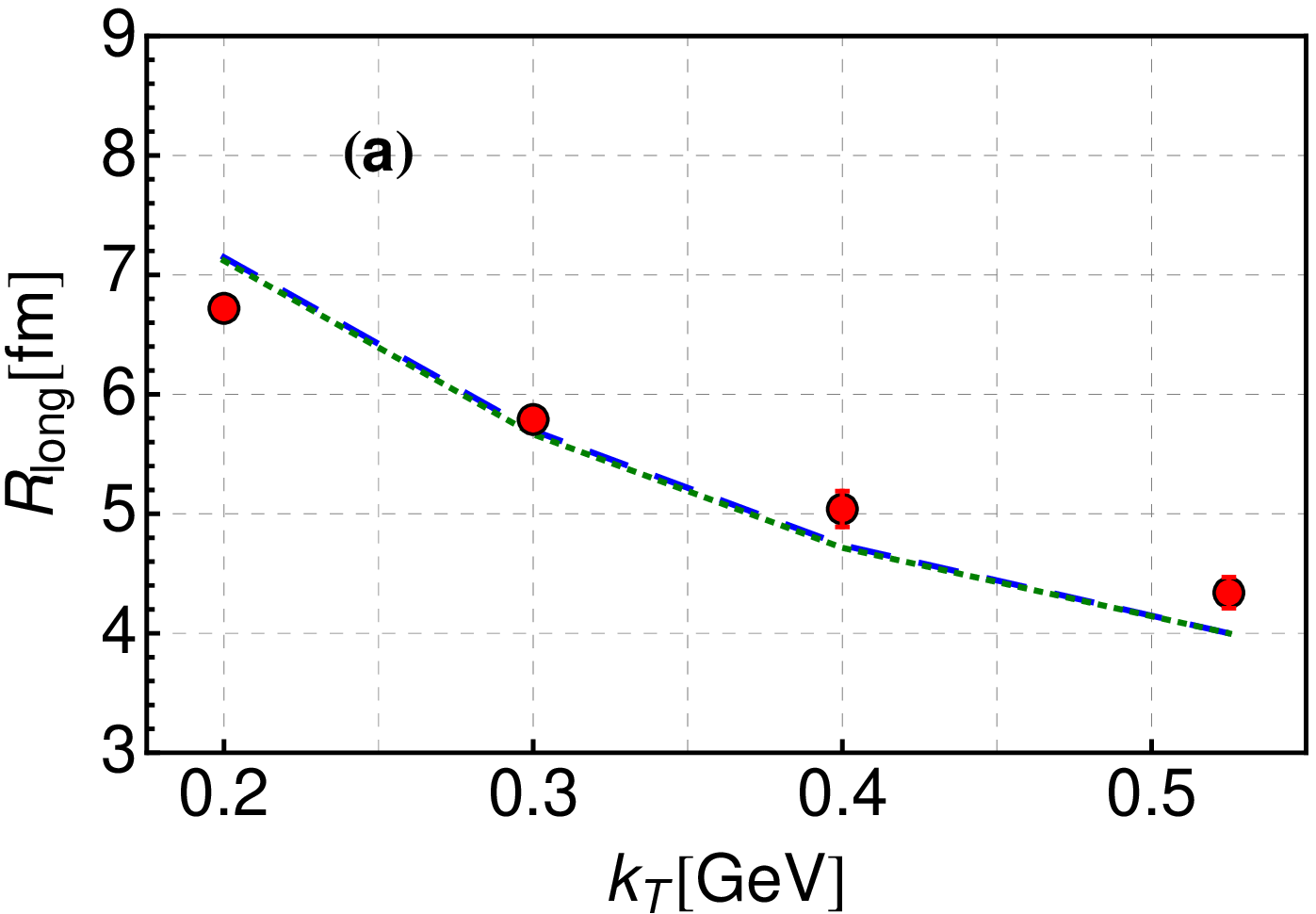}}
\subfigure{\includegraphics[angle=0,width=0.35\textwidth]{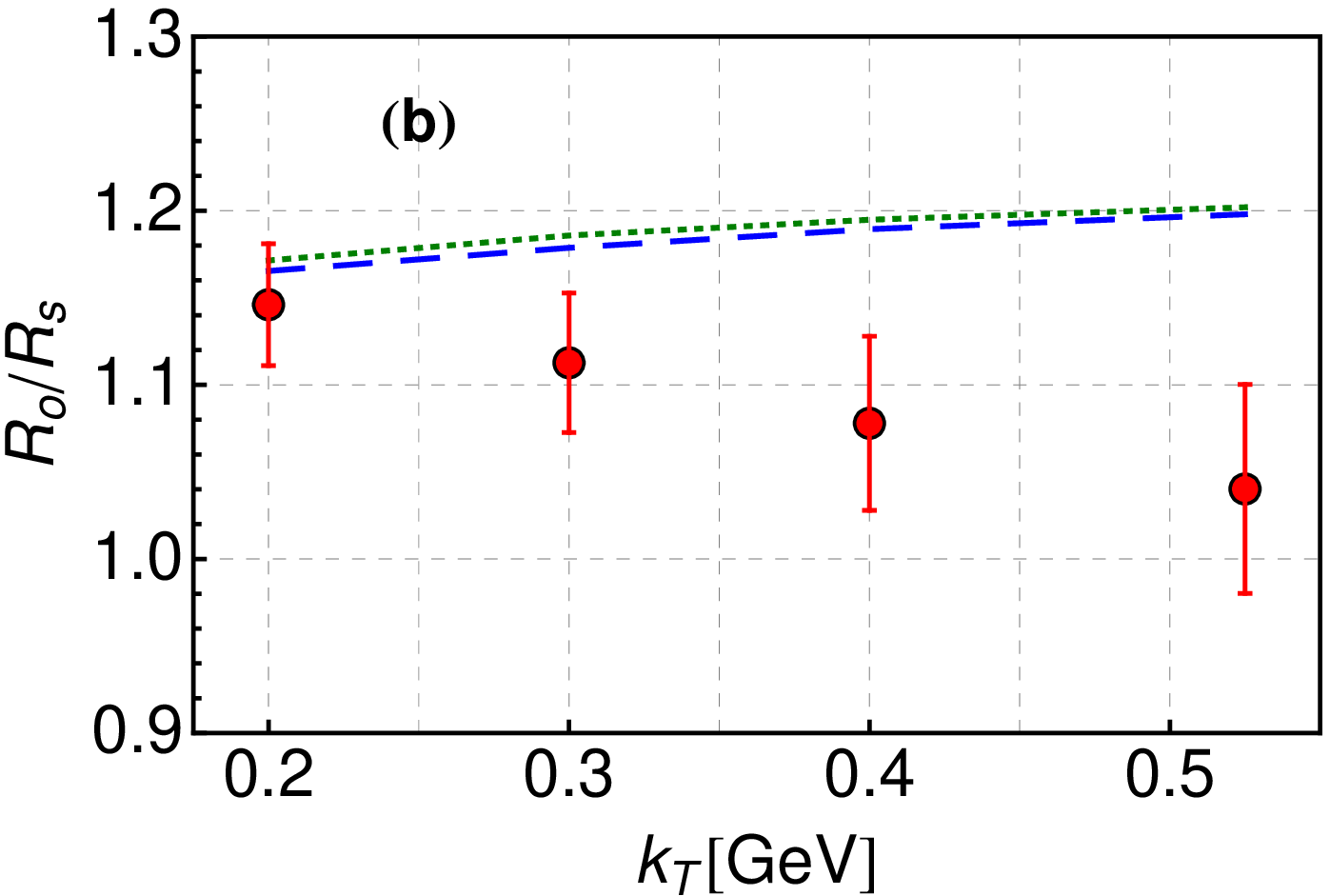}}
\end{center}
\caption{\small (Color online) The same as Fig.~\ref{fig:hbt11} but for the radius $R_{\rm long}$ \textbf{(a)} and the ratio $R_{\rm out}/R_{\rm side}$ \textbf{(b)}. 
}
\label{fig:hbt12}
\end{figure}

\subsubsection{Directed flow}
\label{sect:3Dv1}

\par When one considers non-central collisions of heavy-nuclei, one may observe non-zero directed flow in the region where $\eta_{PS} \neq 0$. This observable is sensitive to both transverse and longitudinal pressures, which makes it an interesting probe for measuring the early local anisotropies in momentum \cite{Bozek:2010aj}. In Figs.~\ref{fig:dirfloweta11}~and~\ref{fig:dirfloweta12} we show the directed flow of charged particles for the two centrality bins, $c=0-5$\% and $c=5-40$\% ($b=6.79$ fm), and for the two time-scale parameters $\tau_{\rm eq}=0.25$ fm and $\tau_{\rm eq}=1.0$ fm. The results are plotted together with the experimental points from STAR \cite{Abelev:2008jga} and PHOBOS \cite{Back:2005pc}. 

For different choices of our model parameters, the results describing $v_1$ are rather stable and consistent with the data. This leads to the conclusion that {\it all kinds of anisotropic stages studied in this work} are consistent with the $v_1$ data. Moreover, the agreement with the data suggests the validity of the idea of a tilted initial source. 

\subsection{HBT correlations}
\label{sect:hbt1}


\par The identical particle interferometry is a useful tool used in the analysis of the system sizes at freeze-out. We calculate the HBT radii using the two-particle method without Coulomb corrections as implemented in {\tt THERMINATOR}~\cite{Kisiel:2005hn,Chojnacki:2011hb}. The details of this procedure have been described in Refs.~\cite{Kisiel:2006is,Kisiel:2008ws}. In Figs.~\ref{fig:hbt11}~and~\ref{fig:hbt12} we present the model results together with the STAR data \cite{Adams:2004yc} for the HBT radii: $R_{\rm out}$, $R_{\rm side}$, and $R_{\rm long}$, all shown as functions of the pair total transverse momentum $k_T$. The model calculations have been done for the time-scale parameter $\tau_{\rm eq}=1.0$ fm, and for the two values of the initial anisotropy: $x_0=100$ (dashed blue lines) and $x_0=0.032$ (dotted green lines). For all considered choices of the parameters, the differences between the data and the model results are smaller than 10\%. We observe that the slopes of the radii are quite well reproduced by our model. At large $k_T$, the side radius,  $R_{\rm side}$, is slightly smaller than that observed in the experiment, and the out radius, $R_{\rm out}$, is slightly larger. This is reflected in the overprediction of the $R_{\rm out}/R_{\rm side}$ ratio by about 20\%. Therefore, the $k_T$ slope of the ratio $R_{\rm out}/R_{\rm side}$ cannot be well reproduced. 

We obtain almost identical results for the two considered values of $x_{0}$. This is expected, since the correct renormalization of the initial energy density results in the similar multiplicity and flow on the freeze-out hypersurface. The similarities in the description of the correlation radii for $x_0=100$ and $x_0=0.032$ mean that the shapes of the freeze-out hypersurfaces are also very much similar. 

\begin{figure}[t]
\begin{center}
\subfigure{\includegraphics[angle=0,width=0.4\textwidth]{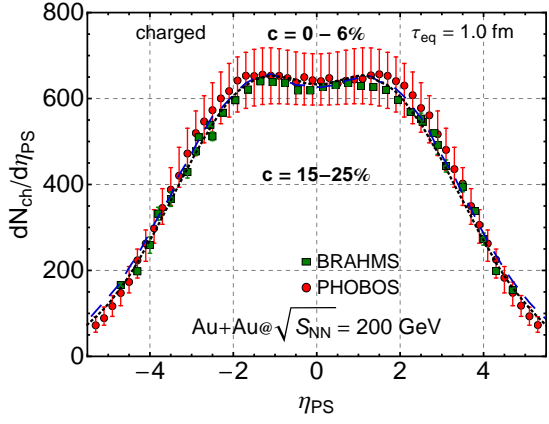}} 
\end{center}
\caption{\small (Color online) The pseudorapidity distributions of charged particles for the case \textbf{i)} (blue dashed line)  and \textbf{ii)} (black dotted line). }
\label{fig:etadistr_RHIC_mp}
\end{figure}

\begin{figure}[t]
\begin{center}
\subfigure{\includegraphics[angle=0,width=0.4\textwidth]{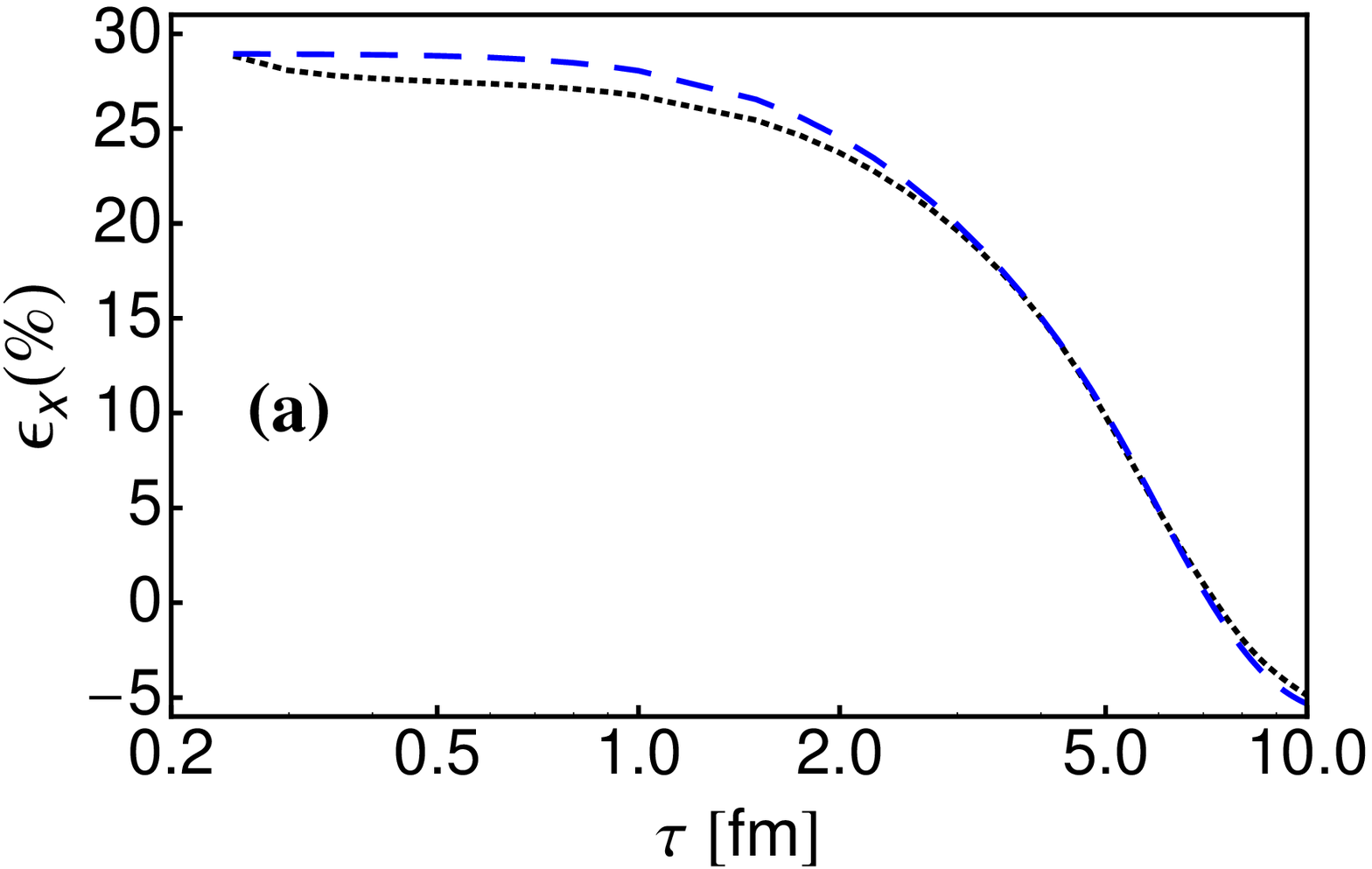}} 
\subfigure{\includegraphics[angle=0,width=0.4\textwidth]{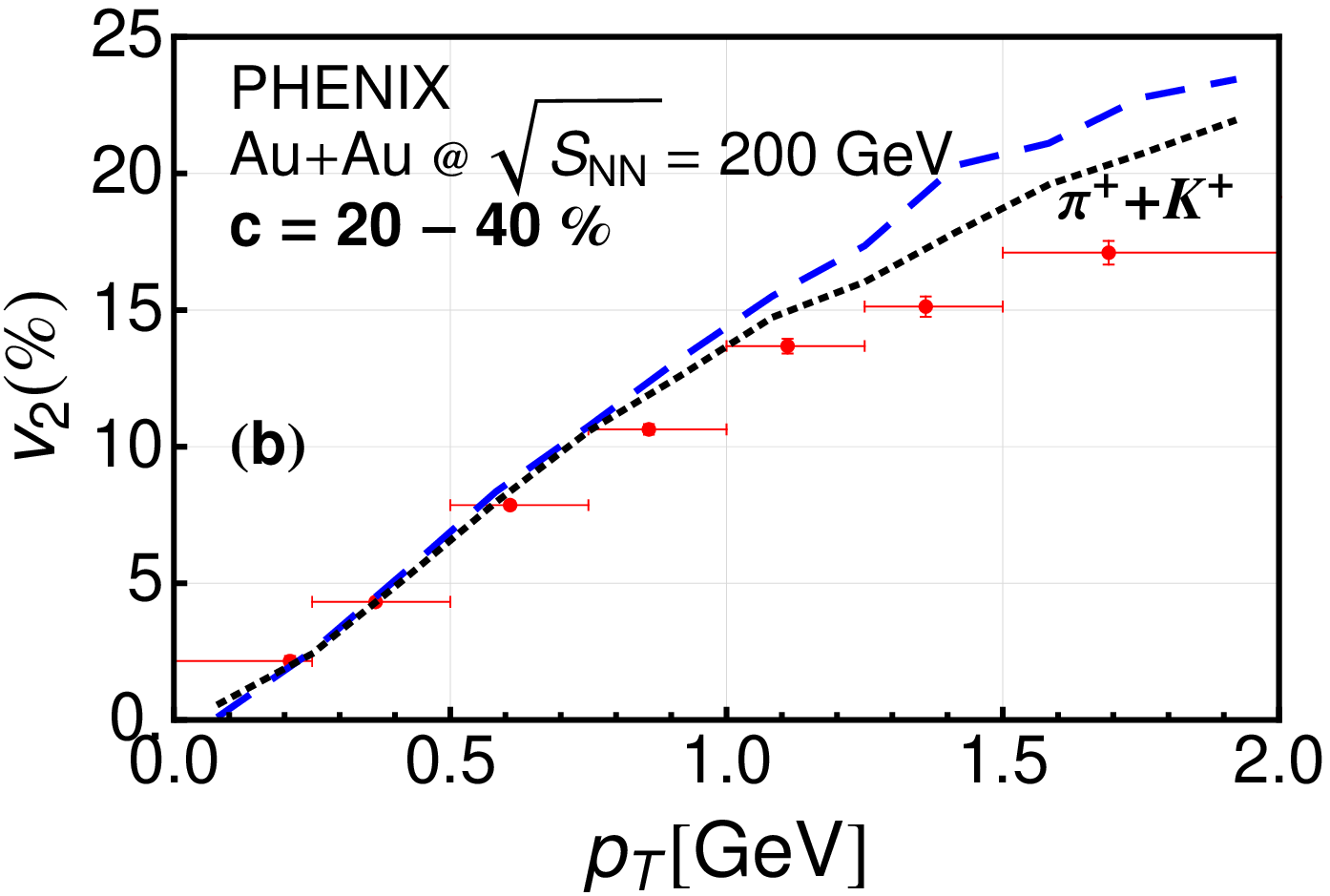}} 
\end{center}
\caption{\small (Color online) \textbf{(a)} Time dependence of the eccentricity of the fireball at $\eta=0$, with the entropy density used as a weight function, for the cases \textbf{i)} (blue dashed line)  and \textbf{ii)} (black dotted line) defined in Sect.~\ref{sect:partspec}. \textbf{(b)} Transverse-momentum dependence of the elliptic flow coefficient $v_2$ of pions and kaons for the cases \textbf{i)} (blue dashed line)  and \textbf{ii)} (black dotted line). }
\label{fig:v2_mp}
\end{figure}

\section{Space dependent initial anisotropy}
\label{sect:partspec}

\par In Ref.~\cite{Hirano:2001eu} it has been suggested that the system may be less equilibrated in forward and backward rapidities (i.e., in the fragmentation regions) than in the midrapidity region. On the other hand, in Ref.~\cite{Bozek:2005eu} it has been argued that the thermalization may depend on the density of participants, since the degree of  thermalization depends on the density of produced particles. Thus, in general, one may expect that $x_{\rm 0}=x_{\rm 0}(\rho(\eta, \bf{x_{\perp}}))$, and $x_{\rm 0} \gg 1$ at the edges of the system. 

Here we check two scenarios: \textbf{i)} a constant initial anisotropy profile $x_{0}=100$ with $\tau_{\rm eq}=1.0$ fm,  $\varepsilon_{\rm i} = 41.8$  $\mathrm{GeV/fm^3}$, $\Delta\eta = 1.5$, and $\sigma_\eta = 1.3$, see Fig.~\ref{fig:etadistr_RHIC_m}, and \textbf{ii)} a space dependent initial anisotropy profile $x_{\rm 0}=x_{\rm 0}(\eta, \bf{x_\perp})$ where
\begin{equation}
x_{\rm 0}(\eta, \bf{x_\perp})=\frac{1}{\tilde{\rho}(\eta, \bf{x_\perp})^2},
  \label{x0prof1}
\end{equation}
with $\tau_{\rm eq}=1.0$ fm,  $\varepsilon_{\rm i} = 73.8$  $\mathrm{GeV/fm^3}$, $\Delta\eta = 1$, and $\sigma_\eta = 1.3$.

Figure~\ref{fig:etadistr_RHIC_mp} shows the pseudorapidity distributions for the cases \textbf{i)} (blue dashed line) and \textbf{ii)} (black dotted line). We observe that the distributions are similar despite different initial anisotropy profiles. Note that in the case \textbf{ii)} $\Delta\eta$ is much smaller, since in the central rapidity region the matter evolves faster in the longitudinal direction than in the case \textbf{i)}. Moreover, moving to forward rapidities the anisotropy grows due to the ansatz (\ref{x0prof1}). Thus, the tails of the pseudorapidity distribution are described more precisely and the results describing the transverse momentum spectra are very similar in the two cases. 

\par Interesting observations may be done for the $v_2$ coefficient at midrapidity. In the case \textbf{i)}, the imposed initial momentum anisotropy is large and constant in the entire fireball. The rescaling of the initial energy density cancels the effect of the larger initial pressure in the transverse direction. Moreover, the eccentricity of the initially produced entropy in the case \textbf{i)}, $\Sigma(x_0,\sigma(\tau_0, \bf{x_\perp}))$, is the same as the eccentricity of the initial entropy density profile. Hence the eccentricity of the fireball transforms into the same momentum anisotropy as in the perfect-fluid case and we obtain similar results for $v_2$. In the case \textbf{ii)} the initial momentum anisotropy is space dependent. In the center of the system it is almost isotropic due to high density of wounded nucleons. The momentum anisotropy is much larger in the regions of low density of wounded nucleons, thus the eccentricity of the initially produced entropy in the case \textbf{ii)} is lower than that in the case \textbf{i)}. It results in a slightly lower eccentricity of the whole fireball and, in consequence, it suppresses slightly the generation of $v_2$, see Fig.~\ref{fig:v2_mp}.

\begin{figure}[t]
\begin{center}
\subfigure{\includegraphics[angle=0,width=0.35\textwidth]{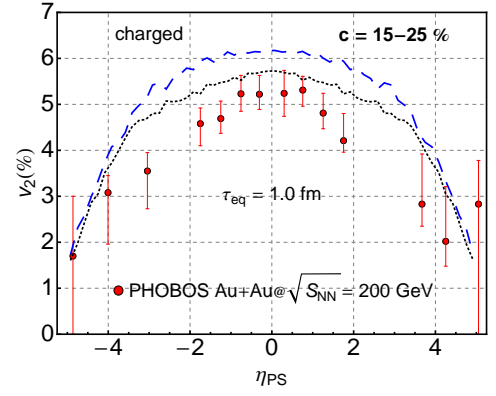}} 
\end{center}
\caption{\small (Color online) The pseudorapidity distributions of the $p_T$-integrated elliptic flow for the cases \textbf{i)} (blue dashed line)  and \textbf{ii)} (black dotted line) defined in Sect.~\ref{sect:partspec}. }
\label{fig:v2eta_mp}
\end{figure}

\begin{figure}[t]
\begin{center}
\subfigure{\includegraphics[angle=0,width=0.35\textwidth]{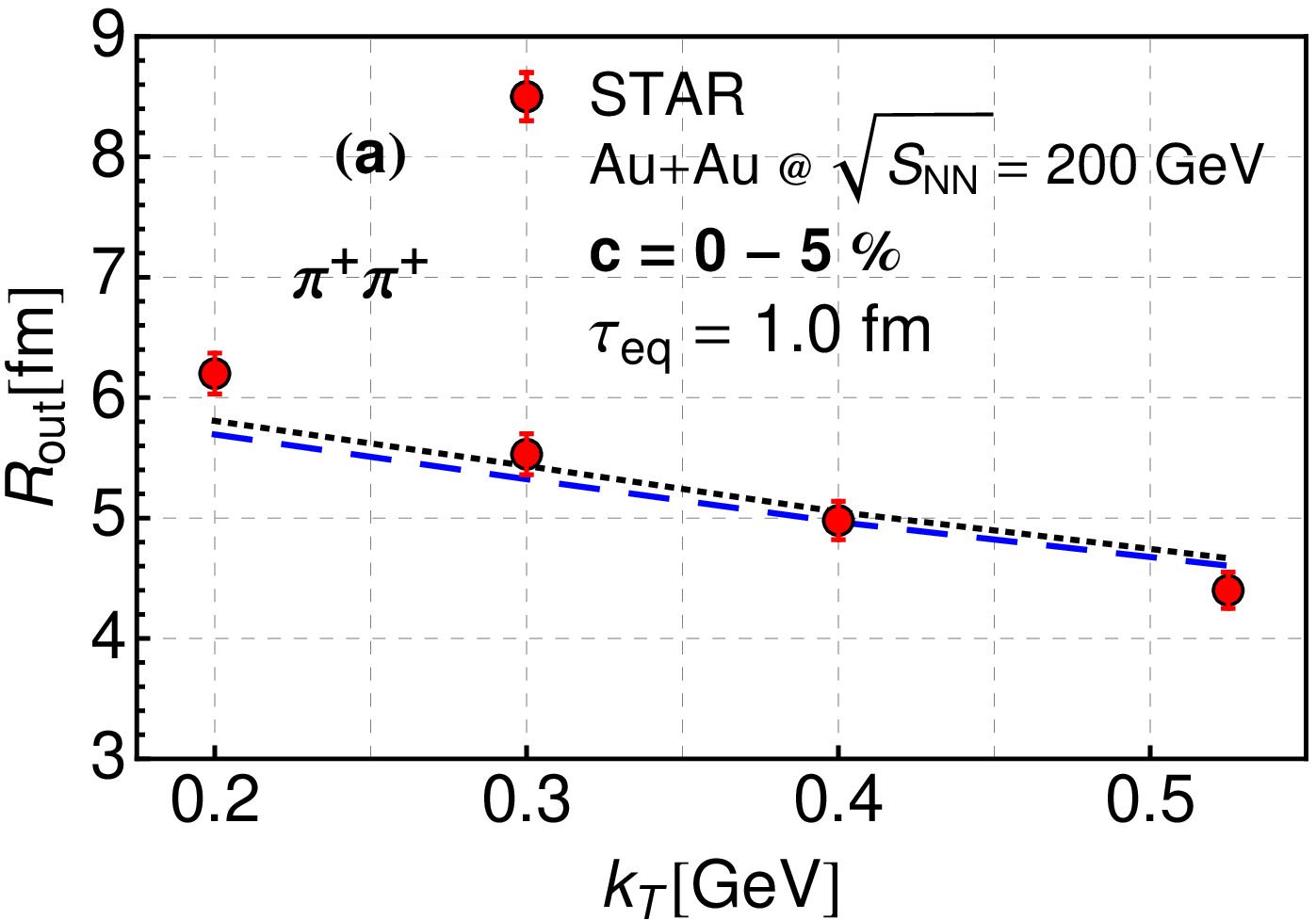}}
\subfigure{\includegraphics[angle=0,width=0.35\textwidth]{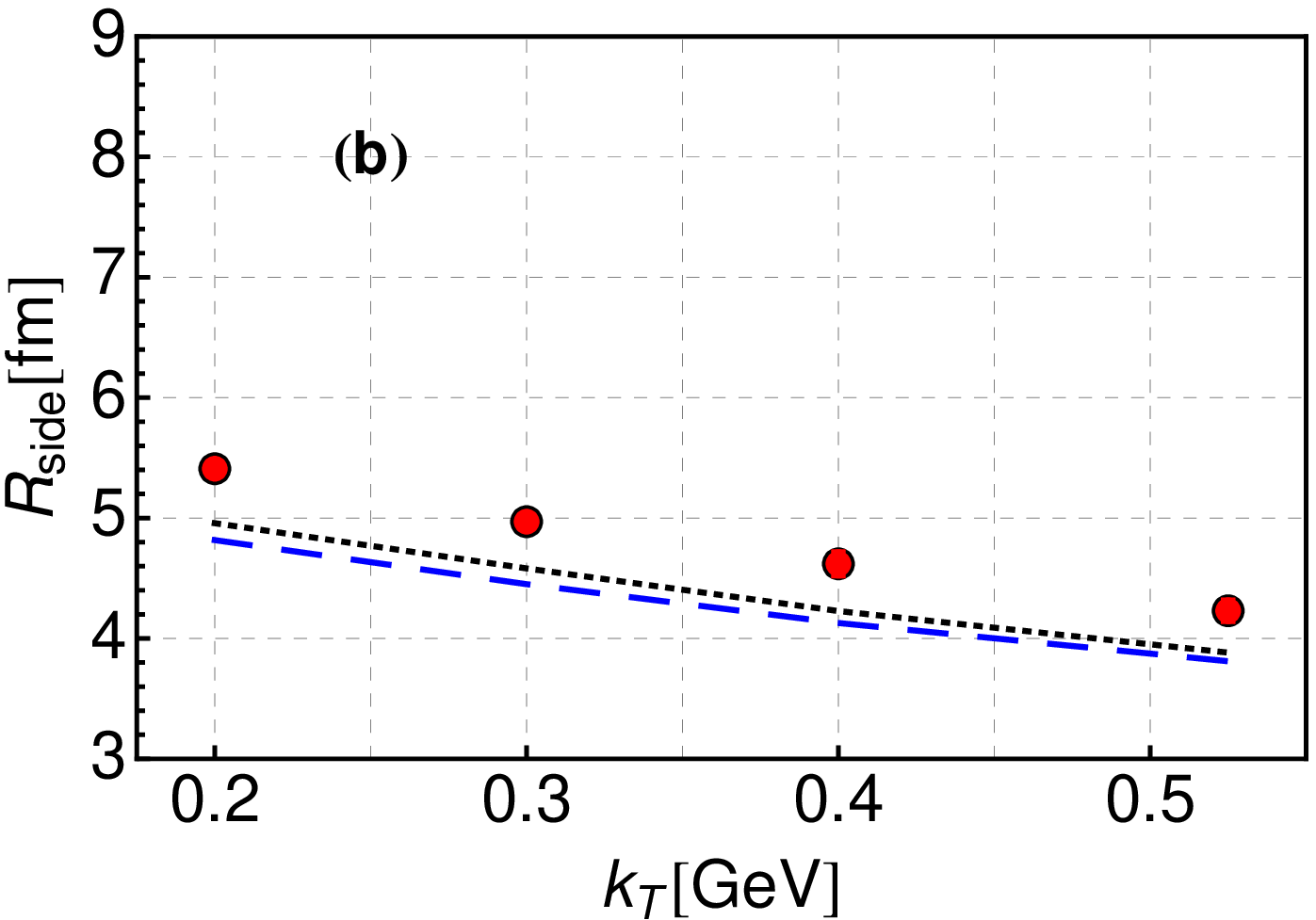}}
\end{center}
\caption{\small (Color online) The HBT correlation radii $R_{\rm out}$ \textbf{(a)} and $R_{\rm side}$ \textbf{(b)} of positive pions shown as functions of total transverse momentum of the pair for the centrality $c=0-5$\%, for the cases \textbf{i)} (blue dashed line) and \textbf{ii)} (black dotted line) defined in Sect.~\ref{sect:partspec}.
}
\label{fig:hbt_mod1}
\end{figure}

\begin{figure}[t]
\begin{center}
\subfigure{\includegraphics[angle=0,width=0.35\textwidth]{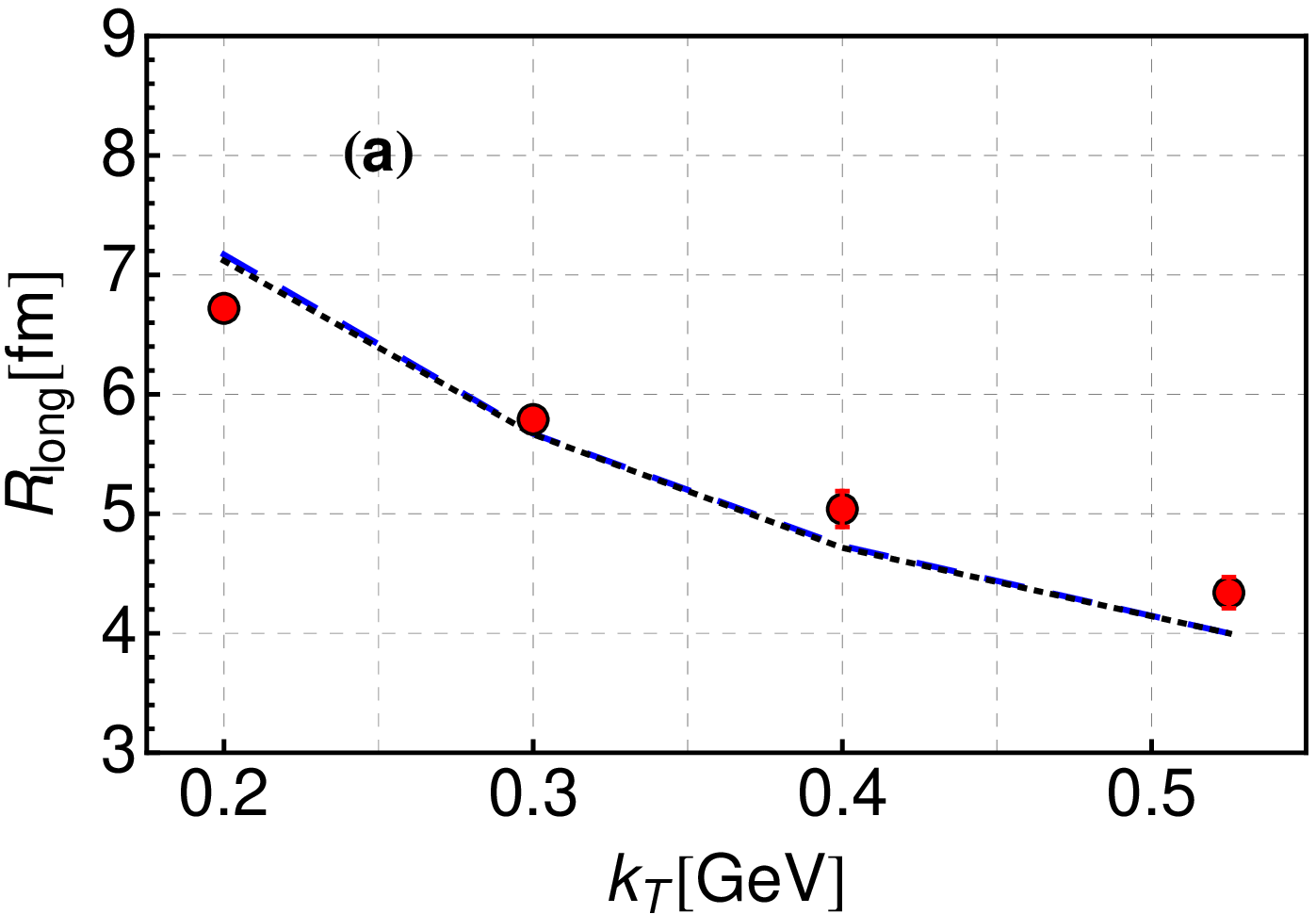}}
\subfigure{\includegraphics[angle=0,width=0.35\textwidth]{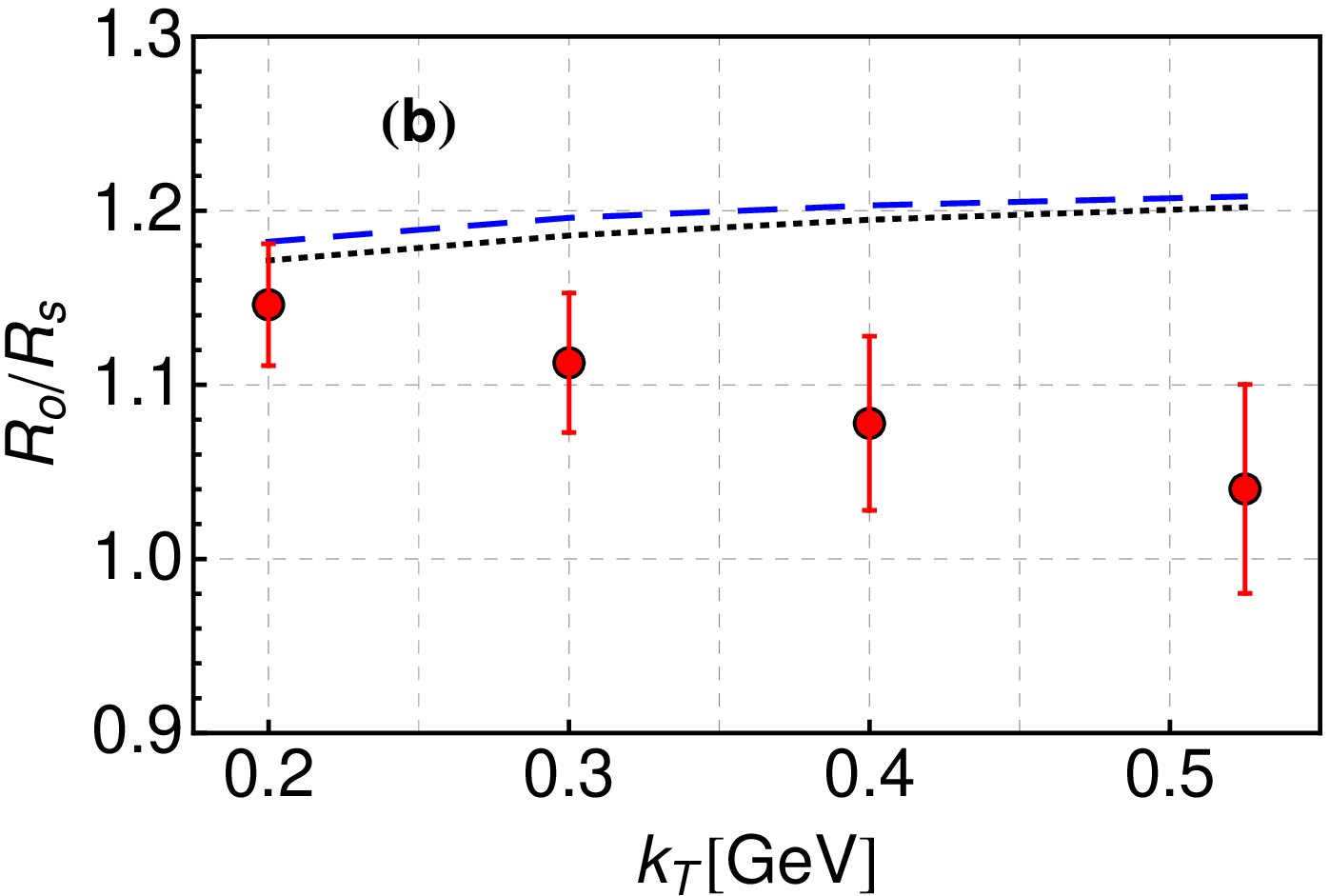}}
\end{center}
\caption{\small (Color online) The same as Fig.~\ref{fig:hbt_mod1} but for the radius $R_{\rm long}$ \textbf{(a)} and the ratio $R_{\rm out}/R_{\rm side}$ \textbf{(b)}.
}
\label{fig:hbt_mod2}
\end{figure}

The effect of lowering of the $v_2$ is clearly seen in Fig.~\ref{fig:v2eta_mp} where we present the pseudorapidity dependence of the $p_{T}$-integrated $v_2$. We observe better agreement of the calculations with the experimental data in the central rapidities in the case \textbf{ii)}. However the model still does not reproduce the peaked shape seen in the data.

The spatial dependence of the initial anisotropy has a small effect on the HBT radii. In Figs.~\ref{fig:hbt_mod1}~and~\ref{fig:hbt_mod2} we present the HBT radii calculated for the case \textbf{i)} (blue dashed line)  and \textbf{ii)} (black dotted line). We observe a small improvement of $R_{\rm side}$ which results in better agreement of $R_{\rm out}/R_{\rm side}$ in the case \textbf{ii)} as compared to the case \textbf{i)}.

\section{Summary and conclusions}
\label{chapter:sum}

In this paper, a recently developed framework of highly-anisotropic and strongly-dissipative hydrodynamics -- \texttt{ADHYDRO} -- has been used in 3+1 dimensions to analyze the space-time evolution of matter produced in ultra-relativistic heavy-ion collisions. The main goal of this analysis was to study possible effects of initial highly-anisotropic stages (in momentum space) on the final soft hadronic observables typically measured in the experiment. 

The study was done in the context of the heavy-ion measurements performed at RHIC (Au+Au collisions at the highest beam energy \mbox{$\sqrt{s_{\rm NN}} = 200$~GeV}). The results of the hydrodynamic models were coupled to the statistical Monte-Carlo model \texttt{THERMINATOR}~\cite{Kisiel:2005hn,Chojnacki:2011hb} which has allowed us to perform systematic study of soft-hadronic observables such as: pseudorapidity distributions, transverse-momentum spectra, directed and elliptic flows, and the HBT radii. The new (3+1)D version of the model allowed us to investigate the pseudorapidity distributions of multiplicity, the transverse-momentum spectra at different values of rapidity, and the pseudorapidity dependence of the $p_T$-integrated elliptic and directed flows.

We have studied systems exhibiting large initial momentum anisotropy corresponding to prolate \mbox{($0 \leq x_0 \ll 1$)} and oblate ($x_0 \gg 1$) configurations. We have also analyzed the space dependent initial momentum anisotropy. The \texttt{ADHYDRO} framework allows us to determine the further evolution of such anisotropic systems and their approach towards local equilibrium. The process of equilibration is controlled by the time-scale (relaxation-time) parameter, for which we used three values: $\tau_{\rm eq}=0.25$ fm, $\tau_{\rm eq}=0.5$ fm, and $\tau_{\rm eq}=1.0$ fm.

We have found that all studied observables are almost insensitive to the initial anisotropic stage provided the initial conditions of the evolution are properly readjusted. This result shows that the early thermalization phenomenon is not required to describe hadronic data, in particular, to reproduce the elliptic flow $v_2$. The complete thermalization of matter (local equilibration) may take place only at the times of about 1-2 fm/c, in agreement with microscopic models.

%

\end{document}